\documentclass[bibyear]{aa}
\usepackage{amsmath,amstext}
\usepackage{natbib,twoopt}
\usepackage[breaklinks=true,colorlinks=true,linkcolor=blue,citecolor=blue,urlcolor=blue]{hyperref}
\usepackage{wrapfig}
\usepackage{amssymb}
\usepackage{wasysym}
\usepackage{float}
\usepackage{ulem}
\usepackage{breakurl}
\usepackage{color}
\usepackage[flushleft]{threeparttable} 

\def\msun{\,{M_\odot}}
\def\spose#1{\hbox to 0pt{#1\hss}}
\def\lta{\mathrel{\spose{\lower 3pt\hbox{$\mathchar"218$}}
     \raise 2.0pt\hbox{$\mathchar"13C$}}}
\def\gta{\mathrel{\spose{\lower 3pt\hbox{$\mathchar"218$}}
     \raise 2.0pt\hbox{$\mathchar"13E$}}}
\def\Mh{{M_{\rm BH}}}

\begin{document}

\newenvironment{tablehere}
  {\def\@@captype{table}}
  {}
\newenvironment{figurehere}
  {\def\@@captype{figure}}
  {}
\makeatother

\titlerunning{Wings of little dots}

\title{Wings of little dots: Exponential broad lines from a stratified BLR}

\author{Piero Madau\inst{1,2}
\and
Roberto Maiolino\inst{3,4,5}
\and
Jan Scholtz\inst{3,4}
\and 
Francesco D’Eugenio\inst{3,4}
}
\institute{Dipartimento di Fisica ``G. Occhialini,'' Università degli Studi di Milano-Bicocca, Piazza della Scienza 3, I-20126 Milano, Italy \and Department of Astronomy \& Astrophysics, University of California, 1156 High Street, Santa Cruz, CA 95064, USA \and
Kavli Institute for Cosmology, University of Cambridge, Madingley Road, Cambridge CB3 0HA, UK 
\and Cavendish Laboratory, University of Cambridge, 19 JJ Thomson Avenue, Cambridge CB3 0HE, UK \and
Department of Physics and Astronomy, University College London, Gower Street, London WC1E 6BT, UK}

\abstract{We investigate the origin of the broad exponential wings seen in the H$\alpha$ profiles of many JWST-discovered little red dots (LRDs) and little blue dots (LBDs). Recent studies show that exponential broad-line profiles are not unique to LRDs, are common in LBDs too, and need not imply that electron scattering is the dominant broadening mechanism in every source. Motivated by our unification picture, in which LRDs are dust-reddened, high-inclination counterparts of compact blue broad-line AGNs, we model the broad Balmer emission with a virialized, radially stratified broad-line region (BLR). In this framework, the observed profile is the luminosity-weighted superposition of clouds spanning a range of radii, and hence virial velocities. We show that a stratified BLR reproduces the exponential-like wings observed in three representative LRDs and one LBD, without requiring electron scattering as the dominant broadening mechanism. For the LBD GS-3073, the same BLR -- calibrated only to H$\alpha$ -- predicts, parameter-free, that He~{\sc ii}~$\lambda4686$ and H$\beta$ should be broader than H$\alpha$, consistent with observations and offering an independent test of BLR ionization stratification. Our results support a picture in which wings and cores encode different physics: the wings arise primarily from virial BLR stratification, while the cores retain imprints of absorption and radiative transfer in dense gas. The fits further suggest that the cloud radial distribution peaks near the dust sublimation radius, with the exponential wings shaped by the inner, line-emitting BLR shells whose higher virial velocities produce the high-velocity tails. This offers a simple physical explanation for the exponential wings of little dots, without invoking exotic new components or scenarios.
}


\maketitle
\section{Introduction}
\label{sec:intro}

The James Webb Space Telescope (JWST) has uncovered a large population of moderate-luminosity broad-line active galactic nuclei (BLAGNs) at $z \gtrsim 4$, powered by accretion onto early massive black holes with inferred masses of order $10^{6}$--$10^{8}\,M_\odot$ \citep[e.g.,][]
{Harikane2023AGN,MaiolinoAGN,Taylor2025_BHMF,Juod2026}, and partitioned by their UV-optical continuum slopes into ``little red dots'' (LRDs) and ``little blue dots'' (LBDs) \citep{Brazzini2026}.
These sources are distinguished by broad Balmer emission, compact morphologies, weak X-ray emission, and, in the case of LRDs, red or V-shaped UV-optical continua with strong Balmer breaks. Their spectra often lack the prominent high-ionization features typical of classical unobscured AGNs, indicating that the physical conditions of the line-emitting and circumnuclear gas differ substantially from those in standard quasar populations \citep[e.g.,][]{Matthee2024,Greene2024,Lambrides2024, Kocevski2025,Wang2025,Hainline2025,Akins2025a,Delvecchio2025,Barro2026,Hviding2025,Maiolino2025,Tang2025,Zucchi2026}.

A major focus of recent work has been the origin of the broad H$\alpha$ line profiles in these systems, and more generally of the broad hydrogen and helium emission lines. \citet{Rusakov2026} argued that, in high-quality LRD spectra, the broad exponential wings are produced primarily by Thomson scattering in compact, Compton-thick ionized cocoons. In this interpretation, the intrinsic line cores are much narrower, implying substantially lower black-hole masses than standard virial estimates. By contrast, \citet{Scholtz2026} showed from a larger sample of 32 JWST AGNs that exponential profiles are not unique to LRDs, are often also present in LBDs, and are not universally preferred over Lorentzian or multicomponent Gaussian descriptions. They further demonstrated that exponential wings can arise naturally from the superposition of virialized BLR clouds spanning a range of radii, without requiring scattering to dominate the line broadening.

At the same time, \citet{Matthee2026} have emphasized that the line cores vary systematically with continuum color and Balmer-break strength: blue sources show relatively narrow central cores, redder systems exhibit P-Cygni-like absorption, and the reddest sources display absorption-dominated cores, while broad exponential wings remain present throughout the sequence. This suggests that the wings and cores may encode different physics, with the former tracing the intrinsic BLR velocity field and the latter the radiative-transfer and absorption effects of dense circumnuclear gas.

In a recent paper, \citet{MadauMaiolino2026} proposed that LRDs are the dust-reddened, high-inclination counterparts of compact blue broad-line AGNs powered by super-Eddington accretion. In that picture, a geometrically thick funnel produces a strongly anisotropic ionizing continuum, while an equatorially concentrated BLR and a modest dusty screen explain the large Balmer equivalent widths, weak high-ionization lines, and red continua of LRDs without invoking a fully enclosing cocoon. 
Motivated by this framework, here we investigate whether the observed exponential H$\alpha$ wings can be reproduced by a virialized, radially stratified BLR and whether the detailed line cores require additional absorption. We first test the model against three absorbed LRDs and then extend the analysis to the unabsorbed LBD GS-3073, assessing whether BLR stratification provides a common origin for the broad wings of both populations. GS-3073 offers a complementary test because, unlike the three LRDs, its H$\alpha$ profile shows no detected Balmer absorption, allowing both broad wings to be fitted, while its detected broad He~{\sc ii}~$\lambda4686$ emission provides an additional probe of ionization-dependent BLR stratification.

\section{The stratified BLR of little dots}

A natural alternative to dominant electron scattering is that broad Balmer profiles are produced by a radially stratified BLR, in which clouds at different radii contribute different characteristic velocities. This interpretation is motivated by the fact that exponential wings are not unique to LRDs, are also found in LBDs, and can arise naturally from the stratification of virialized BLR clouds rather than from a separate scattering cocoon. \citet{Scholtz2026} explicitly argue that exponential wings can emerge from BLR stratification in virial motion, and that stacking or superposing multiple kinematic BLR components tends to produce profiles that are indistinguishable from exponential, even when the individual components are not themselves exponential. Exponential or exponential-like broad-line wings are also not unique to JWST-discovered sources. A classic case is the H$\alpha$ profile of the low-luminosity Seyfert NGC 4395, where symmetric exponential wings were identified by \citet{Laor2006}; more generally, \citet{Kollatschny2013} discussed exponential profiles as one of the standard phenomenological broad-line shapes in classical AGNs.  

The logic of our model is simple. In a super-Eddington funnel geometry, the escaping continuum is highly anisotropic: near-polar observers see a brighter, harder ionizing spectral energy distribution (SED), whereas along equatorial directions self-shadowing suppresses the hardest XUV/soft-X-ray photons and reshapes the continuum \citep{Wang2014,Lupi2024b,Madau2026}. 
In addition, efficient Compton cooling of the hot corona by the intense soft-photon field in the inner accretion flow may intrinsically suppress hard X-ray emission, even in the absence of line-of-sight obscuration \citep{Madau2024,Trinca2026}. If the BLR is concentrated toward the equatorial plane, it is illuminated by a softer, more self-shielded continuum than that seen by low-inclination observers, naturally weakening high-ionization lines while preserving strong Balmer emission. A modest-covering dusty component associated with the outer BLR or a compact circumnuclear obscurer then reddens high-inclination sightlines into the V-shaped continua of LRDs, while lower-inclination views appear as LBDs \citep{MadauMaiolino2026}. In this picture, the observed differences between LRDs and LBDs arise primarily from orientation and line-of-sight processing rather than from intrinsically different engines or evolutionary stages \citep[e.g.,][]{Kido2025,Naidu2025,Pacucci2026}. The model predicts broader and more extended high-EW tails for the Balmer lines in LRDs than in LBDs, owing to the suppression of the observed continuum along dust-intercepted, high-inclination sightlines.

Following \citet{MadauMaiolino2026}, we model the BLR as an equatorially concentrated cloud distribution by specifying the mean number of clouds along a direction of polar angle $i$ as
\begin{equation}
N(i)=N_0\,
\exp\!\left[-\frac{\cos^2 i}{2\sigma_{\rm BLR}^2}\right],
\label{eq:Nlos}
\end{equation}
where $\sigma_{\rm BLR}$ sets the angular thickness of the cloud distribution about the equatorial plane. The corresponding escape probability for direct disk photons is
\begin{equation}
P_{\rm esc}(i)=\exp\!\big[-N(i)\big].
\label{eq:Pof_i}
\end{equation}
We fix the normalization $N_0$ by requiring that the angle-averaged probability of intercepting at least one cloud equals the global BLR covering factor,
\begin{equation}
C_{\rm BLR}=\int_{0}^{\pi/2}\Big[1-P_{\rm esc}(i)\Big]\sin i\,{\rm d}i,
\label{eq:Cblr_norm}
\end{equation}
where symmetry about the mid-plane has been assumed. For any chosen pair of values $(\sigma_{\rm BLR},C_{\rm BLR})$, equation~(\ref{eq:Cblr_norm}) is solved for $N_0$. The corresponding normalized angular weighting of BLR clouds is then
\begin{equation}
p(i)=
\frac{[1-P_{\rm esc}(i)]\sin i}
{\int_0^{\pi/2}[1-P_{\rm esc}(i')]\,\sin i'\,{\rm d}i'}.
\label{eq:p_i_blr}
\end{equation}
In our picture, the BLR extends over a range of radii interior to the dust sublimation boundary. This is physically motivated by the fact that, beyond the sublimation radius, surviving dust dominates over gas absorption of ionizing photons, so that only a small fraction of the ionizing continuum is available for gas ionization and line emission. Nebular emission is therefore strongly suppressed there, in addition to being affected by dust extinction \citep{NetzerLaor1993}. We write the shell-integrated radial cloud distribution as
\begin{equation}
\frac{dN}{dr} \propto r^{\alpha-1}\exp\!\left(-\frac{r}{r_{\rm sub}}\right),
\label{eq:dNdr}
\end{equation}
where $r_{\rm sub}$ is the dust sublimation radius \citep{Barvainis1987}, 
\begin{equation}
r_{\rm sub}=1.4\,
\left(\frac{L_{\rm UV}}{10^{45.5}\,{\rm erg\,s^{-1}}}\right)^{1/2}
\left(\frac{T_{\rm sub}}{1200\,{\rm K}}\right)^{-2.8}\,{\rm pc},
\label{eq:barvainis}
\end{equation}
which we take to define the characteristic outer scale of the BLR, and the slope $\alpha$ controls the relative weighting of inner and outer BLR clouds. Since the emergent H$\alpha$ line power of a cloud, $P_{\rm H\alpha}^c(r,i)$, depends on both radius and BLR angle, the line profile is written as
\begin{equation}
\frac{dL}{dv}(v)=
\int dr\,\frac{dN}{dr}
\int di\,p(i)\,
P_{\rm H\alpha}^c(r,i)\,
G\!\left[v;\sigma(r)\right],
\label{eq:dLdv}
\end{equation}
where $G(v;\sigma)$ is the Gaussian kernel describing the effective one-dimensional distribution of cloud bulk velocities at fixed radius $r$. Here $i$ denotes the polar angle of an individual BLR cloud, and the $p(i)$-weighted integral accounts for the anisotropic illumination of clouds at different angles from the disk axis under our assumption of isotropically re-emitted line radiation; it is therefore distinct from the observer's own line-of-sight inclination $i_{\rm obs}$, which instead enters only through the directly viewed continuum in the equivalent-width calculation of Section~\ref{sec:summary}. Because the incident ionizing rate $Q_{\rm HI}(i)$ is anisotropic, $p(i)$ weights a mixture of angle-dependent radial emissivity distributions. Thus, both $\sigma_{\rm BLR}$ and, through the nonlinear dependence of $p(i)$ on $N_0$, $C_{\rm BLR}$ can influence the predicted wing shape, although the dependence on $C_{\rm BLR}$ becomes weak at small covering factors. We nevertheless fix both at representative values (Section~\ref{sec:lrds}) and vary only $\alpha$ and $f_{\rm vir}$ when fitting the observed profiles. The same fixed values of $C_{\rm BLR}$ and $\sigma_{\rm BLR}$ are used to predict how the total H$\alpha$ EW varies with $i_{\rm obs}$ in Section~\ref{sec:summary}.

Assuming gravitationally dominated motion, the characteristic Keplerian velocity is
\begin{equation}
v_K(r)=\left(\frac{GM_{\rm BH}}{r}\right)^{1/2}.
\end{equation}
In the phenomenological fits considered here, we parameterize the local line-of-sight velocity distribution as a Gaussian with dispersion
\begin{equation}
\sigma(r)=f_{\rm vir}\,v_K(r),
\end{equation}
where the dimensionless velocity-scale parameter \(f_{\rm vir}\) absorbs geometric and kinematic factors, as well as source-to-source differences in black-hole mass relative to the adopted value. Because the profile width scales as \(f_{\rm vir}M_{\rm BH}^{1/2}\), these effects are degenerate in the present model. Any inclination dependence is likewise absorbed into \(f_{\rm vir}\); if an ordered in-plane component were modeled explicitly, its projected velocity would scale as \(\sin i_{\rm obs}\).

The inner BLR therefore contributes the high-velocity tails, while larger radii supply the low-velocity core. In this way, a continuous radial distribution of virialized clouds can generate broad exponential-like wings without requiring scattering to dominate the profile, as detailed in the next section. 
This interpretation also naturally allows different emission lines to have different profiles. As we shall see, higher-ionization lines are weighted toward smaller radii and therefore acquire broader profiles than the Balmer lines. Such line-dependent widths are not naturally expected if a single scattering medium dominates the broadening, but arise straightforwardly in a stratified BLR. More generally, BLR stratification is already a standard ingredient of AGN physics, supported by reverberation-mapping studies showing that higher-ionization lines are emitted closer to the black hole than the Balmer lines \citep[e.g.,][]{Peterson1993,Pancoast2014,Grier2017,Fausnaugh2017,Netzer2020}. Our model therefore does not invoke a new or exotic component, but rather applies a structure already known to be present in AGNs.

\begin{figure*}[!htb]
\centering
\includegraphics[width=\hsize,trim=0 0 0 0,clip]{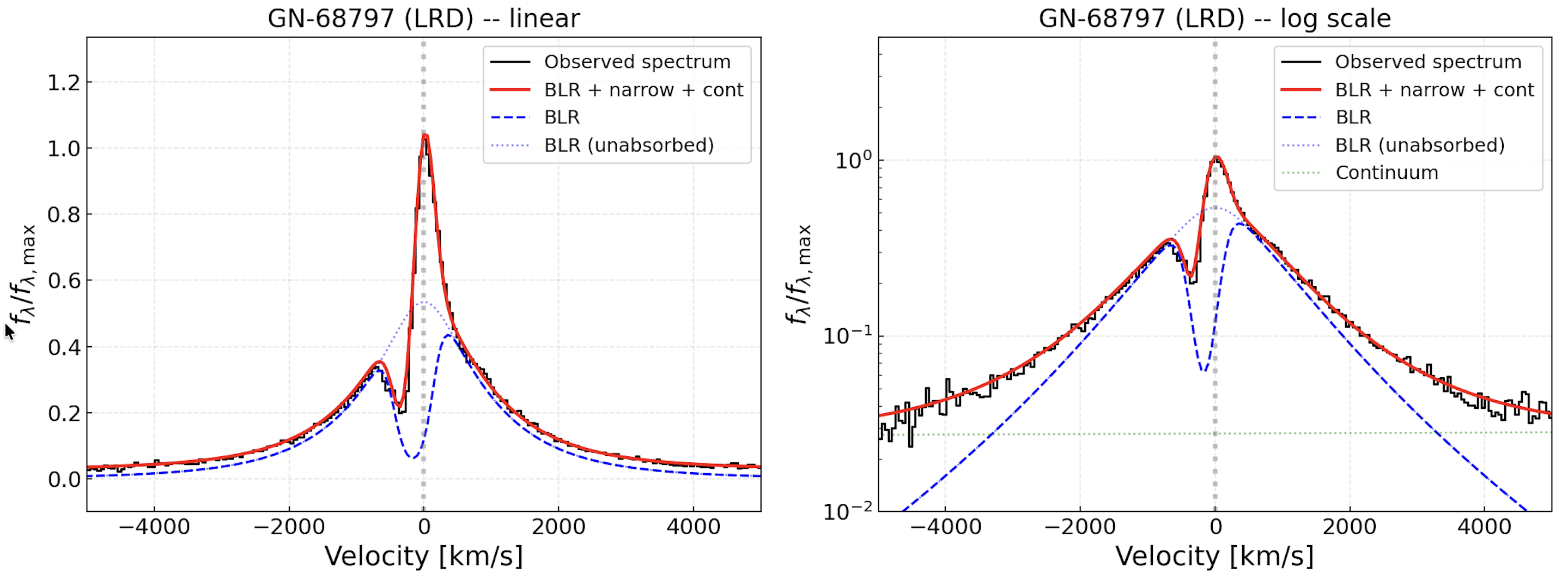}
\caption{Best-fit stratified-BLR model for the LRD GN-68797. Left: linear-scale view of the observed H$\alpha$ profile (black) and best-fit model (red), obtained by fitting only the red wing of the line. The blue dashed curve shows the absorbed broad stratified BLR component, while the blue dotted curve shows the corresponding unabsorbed BLR profile; the difference between the two illustrates the effect of the fixed absorption profile adopted from the published line-profile decomposition
\citep{Scholtz2026}. Right: the same comparison on a logarithmic scale, highlighting the extended, nearly exponential wings. The fit is obtained for a reduced chi-square of $\chi^2/{\rm dof}=1.8$, with $\alpha=1.75$ and $f_{\rm vir}=1.2$, assuming $r_{\rm sub}=7.2\times10^{18}\,{\rm cm}$. The model reproduces the broad wings and the overall profile shape, supporting an interpretation in which the extended wings arise from the superposition of virialized clouds spanning a range of radii, while the suppressed core is shaped by line-of-sight absorption and the fixed narrow H$\alpha$ component.
}
\label{fig:68797}
\end{figure*}

\section{Line profiles of LRDs}
\label{sec:lrds}

To illustrate the implications of our stratified BLR framework, we construct detailed line-profile models for three representative JWST broad-line AGNs: GN-68797, GN-9771, and GS-13971. All three are LRDs with deep NIRSpec grating spectroscopy and are included in the recent line-profile analysis of \citet{Scholtz2026}, where they are classified as absorbed LRDs. Their spectroscopic redshifts are $z=5.04$, $5.53$, and $5.48$, respectively. In the \citet{Matthee2026} sample, the total H$\alpha$ rest-frame equivalent widths inferred from the PRISM spectra are $1306\,$\AA\ for GN-68797, $1713\,$\AA\ for GN-9771, and $894\,$\AA\ for GS-13971, confirming that all three are strong Balmer emitters within the broader JWST broad-line population.

We generated grids of photoionization models with the C23 release of {\sc Cloudy} \citep{Chatzikos2023}, tailored to typical BLR conditions with hydrogen density $n_{\rm H}=10^{10}\,{\rm cm^{-3}}$ and column density $N_{\rm H}=10^{23}\,{\rm cm^{-2}}$. We adopted a metallicity $Z=0.1\,Z_\odot$, representative of galaxies at the redshifts where LRDs (and LBDs) are most commonly found,\footnote{While the H$\alpha$ emissivity is less sensitive to metallicity than metal-line diagnostics, changing $Z$ modifies the thermal balance, ionization structure, and diffuse continuum of the BLR gas, and therefore can affect both the H$\alpha$ line power and its EW at a moderate level.} and explored a grid of ionization parameters $-4 \le \log U \le +1.5$ in steps of 0.5 dex. For the line-profile calculations we adopt a fiducial central engine with $\Mh=10^{7.5}\msun$ and $\dot m=32$, thereby fixing the angle-dependent ionizing photon rate $Q_{\rm HI}(i)$ incident on BLR clouds at polar angle $i$. Each value of \(U\) maps to a characteristic BLR radius through
\begin{equation}
r(U,i)=\left[\frac{Q_{\rm HI}(i)}
{4\pi c\,n_{\rm H}\,U}\right]^{1/2}.
\end{equation}
We use this mapping to assign the Cloudy line intensity calculated at each \(U\) to the corresponding radius. The profile integral is then evaluated directly in radial space, with the cloud population weighted by \(dN/dr\).

For the profile fits, we normalize both the observed spectrum and the model to the peak flux density of the observed total H$\alpha$ profile. 
The BLR covering factor \(C_{\rm BLR}\) and angular thickness \(\sigma_{\rm BLR}\) are fixed to representative values following the equivalent-width analysis of \citet{MadauMaiolino2026}. The radial stratification parameter \(\alpha\) and the dimensionless velocity-scale parameter \(f_{\rm vir}\) are varied in the fits. Because these parameters are strongly degenerate, unconstrained minimization can drive \(f_{\rm vir}\) far from unity for only a marginal reduction in \(\chi^2\). We therefore restrict \(f_{\rm vir}\) to within 40 per cent of unity (\(0.6\leq f_{\rm vir}\leq1.4\)), selecting solutions whose characteristic velocities remain comparable to the Keplerian value adopted in the model. For smooth profile construction, the {\sc Cloudy} H$\alpha$ line powers are log-linearly interpolated across the sampled $\log U$ grid. In GN-68797 and GS-13971, the absorption profile, narrow H$\alpha$ component, and continuum are fixed to the published line-profile decomposition of \citet{Scholtz2026}. For GN-9771, we adopt a two-step procedure: we first fit the full profile with the stratified BLR multiplied by a free absorption component and a free narrow Gaussian, then fix the resulting narrow and absorption profiles and refit only the red wing to determine the final BLR parameters.
Because in all three objects the blue side of the line is affected by absorption, we fit only the red wing over the velocity interval $300 < v < 4000\,{\rm km\,s^{-1}}$. We exclude the innermost $|v|<300\,{\rm km\,s^{-1}}$ around line center because that part of the profile is most sensitive to uncertainties in the systemic velocity, the fixed narrow-line decomposition, and absorption-related core structure. This allows the fit to focus on the velocity range most directly tracing the stratified BLR responsible for the extended wings. We then add the stratified broad-line component and compare the resulting normalized model profiles directly to the data.

Figure \ref{fig:68797} shows the best-fit stratified-BLR model for the LRD GN-68797. Despite the simplicity of the model, the agreement with the data is excellent: the broad exponential-like wings, the suppressed core, and the overall shape of the observed profile are all reproduced well. Fitting the red wing only, we obtain a best-fit with reduced chi-square of $\chi^2/{\rm dof}=1.8$, radial-slope parameter $\alpha=1.75$, and effective virial factor $f_{\rm vir}=1.2$, adopting a sublimation radius of $r_{\rm sub}=7.2\times10^{18}\,{\rm cm}$. Since the cloud distribution per logarithmic radial interval peaks at $r=\alpha r_{\rm sub}$, this places the bulk of the cloud population moderately outside the dust-sublimation radius, rather than at or dominated by the innermost BLR. At the same time, inner clouds still make a disproportionate contribution to the highest-velocity tails, since the local velocity width scales as $v_K\propto r^{-1/2}$. The broad, nearly exponential wings therefore arise naturally from the superposition of virialized clouds spanning a range of radii, rather than from a single characteristic emitting radius.

\begin{figure*}[!htb]
\centering
\includegraphics[width=\hsize,trim=0 0 0 0,clip]{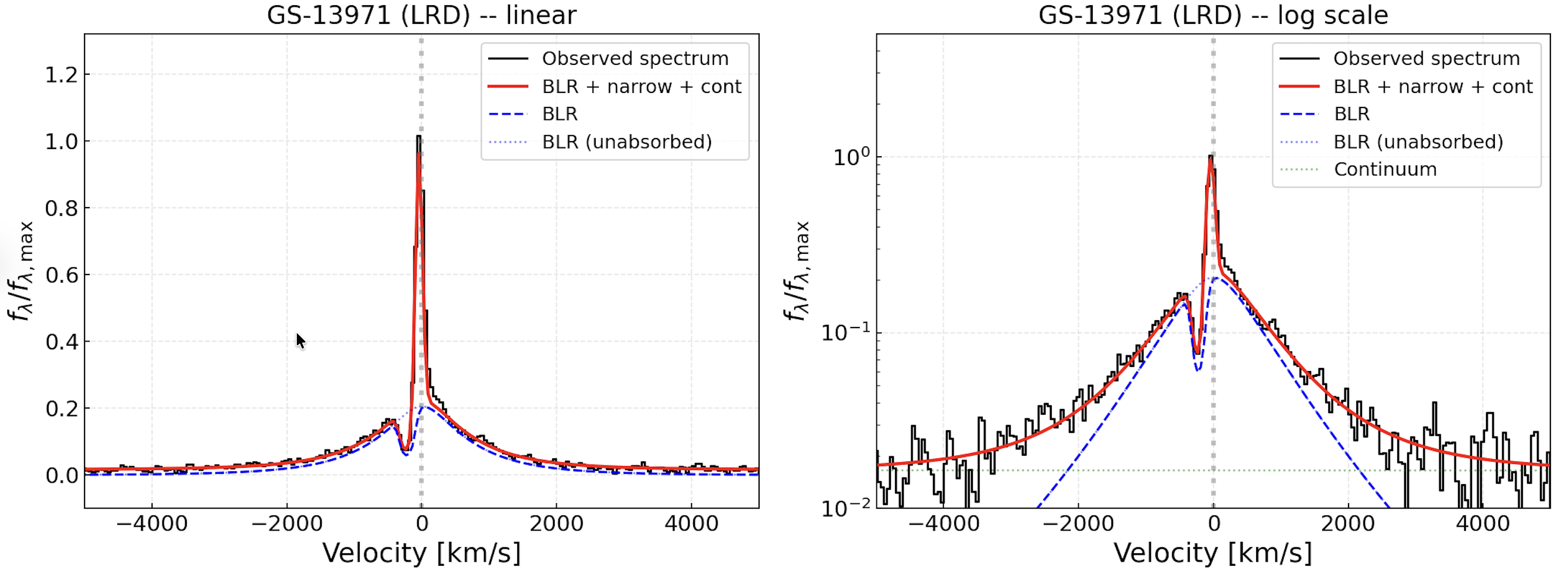}
\caption{Best-fit stratified-BLR model for the LRD GS-13971. Left: linear-scale view of the observed H$\alpha$ profile (black) and best-fit model (red), again fitting only the red wing of the line. The blue dashed curve shows the absorbed broad stratified BLR component and the blue dotted curve the corresponding unabsorbed BLR profile, illustrating the effect of the fixed absorption component adopted from the published decomposition. Right: the same comparison on a logarithmic scale, emphasizing the extended wings. The best fit is obtained for a reduced chi-square of $\chi^2/{\rm dof}=1.0$, with $\alpha=1.75$ and $f_{\rm vir}=0.94$, adopting the same fiducial sublimation radius. 
}
\label{fig:13971}
\end{figure*}

\begin{figure*}[!htb]
\centering
\includegraphics[width=\hsize,trim=0 0 0 0,clip]{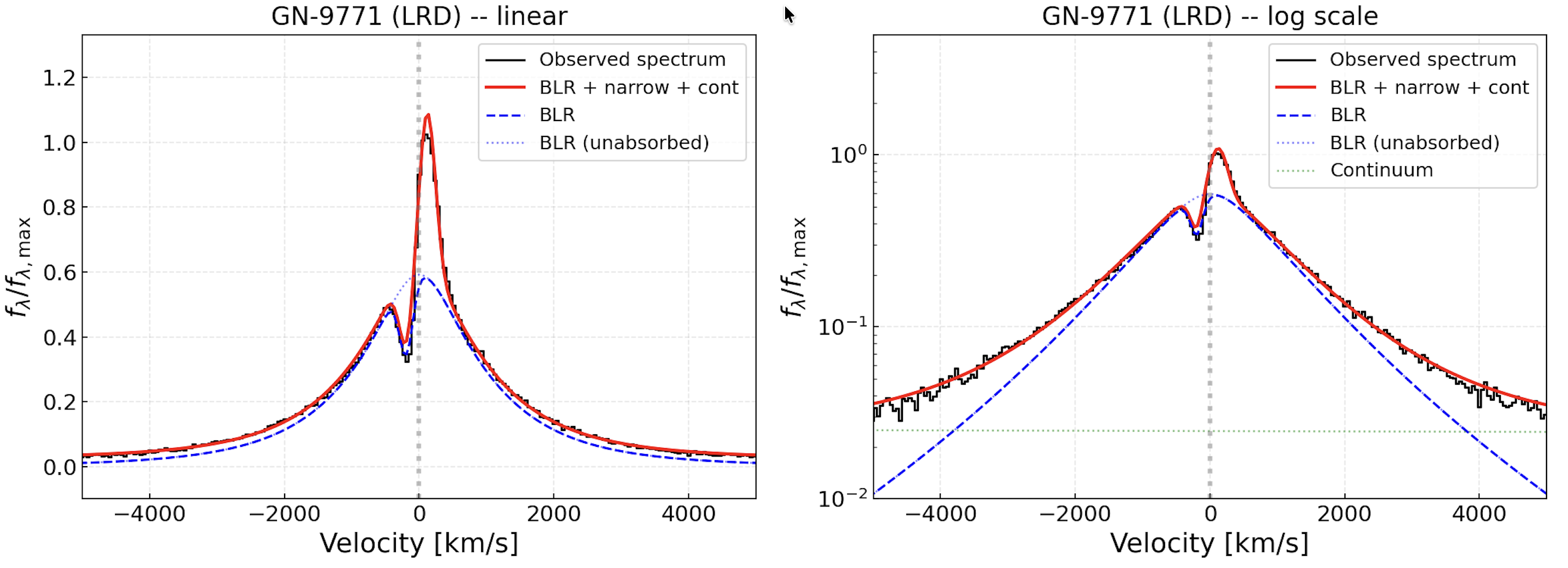}
\caption{Best-fit stratified-BLR model for the LRD GN-9771. Left: linear-scale view of the observed H$\alpha$ profile (black) and best-fit model (red), again fitting only the red wing of the line. The blue dashed curve shows the absorbed broad stratified BLR component and the blue dotted curve the corresponding unabsorbed BLR profile, illustrating the effect of the absorption component determined from the full-profile fit (see text). Right: the same comparison on a logarithmic scale, emphasizing the extended wings. The best fit is obtained for a reduced chi-square of $\chi^2/{\rm dof}=1.4$, with $\alpha=1.75$ and $f_{\rm vir}=1.3$, adopting the same fiducial sublimation radius. As for GN-68797 and GS-13971, the model provides an excellent description of the broad, nearly exponential wings, indicating that a radially stratified BLR weighted toward large radii can account naturally for the observed line profile.
}
\label{fig:9771}
\end{figure*}

This interpretation is qualitatively consistent with the stratified BLR picture inferred in low-redshift AGNs from reverberation mapping and velocity-resolved broad-line studies, which indicate that different parts of the line profile probe gas spanning a range of radii and exhibiting distinct kinematic signatures, including virialized motion, inflow, and outflow \citep[e.g.,][]{Peterson1993,Pancoast2014,Grier2017,Netzer2020}, and with the idea that the low-ionization BLR is associated with the region where the outer accretion disk merges into the inner edge of the dusty torus \citep{Goad2012}.
The parameter $f_{\rm vir}=1.2$ indicates that the effective one-dimensional velocity width required by the fit is somewhat larger than the fiducial virial scaling, absorbing both geometric and kinematic effects as well as any mismatch between the adopted fiducial black-hole mass, $\Mh=10^{7.5}\,M_\odot$, and the true mass of the source. 

Similar results are obtained for the LRDs GS-13971 and GN-9771, whose line-profile decompositions are shown in Figures~\ref{fig:13971} and \ref{fig:9771}. In both cases the agreement between the data and the stratified BLR model is excellent. For GS-13971 we obtain a best fit with reduced chi-square $\chi^2/{\rm dof}=1.0$, $\alpha=1.75$, and $f_{\rm vir}=0.94$, while for GN-9771 we find $\chi^2/{\rm dof}=1.4$, $\alpha=1.75$, and $f_{\rm vir}=1.3$. As in GN-68797, the model reproduces the extended, nearly exponential wings of the other sources. The best-fit radial slopes for all three LRDs, while nearly identical, are degenerate: the $\chi^2$ surface is nearly flat over a wide range of $\alpha$ (GS-13971), or exhibits a shallow secondary minimum tied with a lower value (GN-68797, GN-9771). As for GN-68797, most clouds again reside at radii of order the dust-sublimation scale, with the innermost clouds still dominating the high-velocity tails. Thus, the comparable broad-line morphologies arise from a similar balance between outer-BLR weighting and inner-BLR kinematic broadening, without requiring a universal radial slope.

\begin{figure*}[!htb]
\centering
\includegraphics[width=\hsize,trim=0 0 0 0,clip]{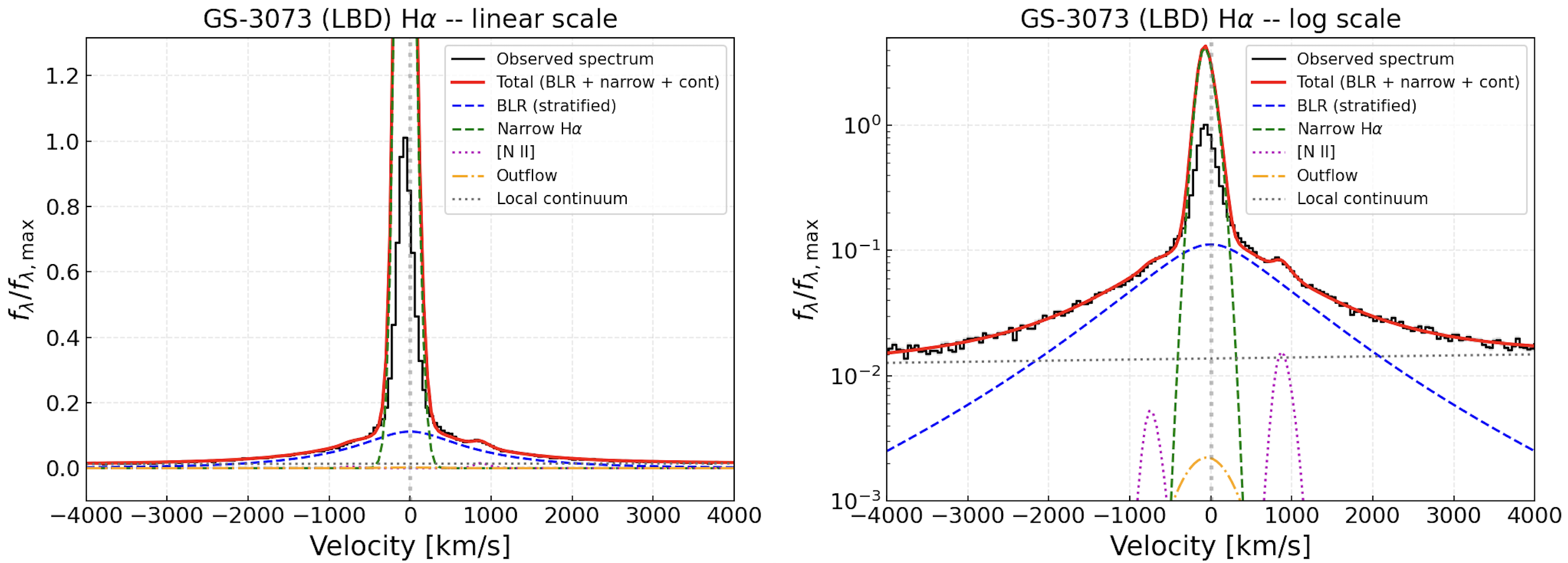}
\caption{Best-fit stratified-BLR model for the LBD GS-3073. Left: linear-scale view of the observed H$\alpha$ profile (black) and best-fit model (red), fitting both wings of the line. The stratified-BLR component is shown together with the refitted narrow H$\alpha$, [N~{\sc ii}], and outflow components; their sum gives the total model. Right: the same comparison on a logarithmic scale, emphasizing the extended wings. For the adopted solution, obtained using the prior on \(\alpha\) motivated by the three LRDs, we find \(\chi^2/{\rm dof}=4.1\), \(\alpha=1.75\), and \(f_{\rm vir}=1.1\), adopting the same fiducial sublimation radius. The model captures the broad-wing morphology, although statistically significant residuals remain, likely associated with the complex narrow-line and outflow components.}
\label{fig:3073}
\end{figure*}

\section{The LBD GS-3073}
\label{sec:gs3073}

We extend our analysis to GS-3073, the ``Blue Rosetta,'' a compact LBD at \(z=5.55\) originally presented by \citet{Ubler2023} and subsequently studied in detail by \citet{Brazzini2026} and \citet{Scholtz2026}. Unlike the absorbed LRDs considered above, GS-3073 has a blue optical and UV continuum and shows no detected Balmer absorption, but its broad H\(\alpha\) line exhibits prominent non-Gaussian wings. \citet{Scholtz2026} identify an exponential profile as the preferred empirical description of H\(\alpha\), making GS-3073 a useful test of whether the same stratified-BLR interpretation extends beyond the LRD population. More importantly, the detection of broad He{\sc ii} \(\lambda4686\) provides a direct test of ionization stratification: if the higher-ionization line originates closer to the black hole, it should exhibit larger characteristic velocities than H\(\alpha\).

For GS-3073 we adopt the same stratified-BLR framework, embedding it in the fuller multi-component decomposition of the H\(\alpha\) complex presented by \citet{Scholtz2026} and \citet{Brazzini2026}. Their model resolves H\(\alpha\) into a narrow component, the [N~{\sc ii}] \(\lambda\lambda6548,6583\) doublet (fixed to the theoretical 3:1 ratio), a blueshifted outflow component, and a broad component described empirically by an exponential, scattering-wing profile; every component except the broad one is a simple Gaussian. Rather than adopt the published narrow/[N~{\sc ii}]/outflow amplitudes as fixed, we retain their centroids and widths but refit their amplitudes as free linear scale factors simultaneously with the stratified BLR component, so that any mismatch between our flux calibration and the published decomposition is absorbed by these nuisance parameters rather than propagating into the wing residuals that constrain the BLR fit. The empirical broad-H\(\alpha\) exponential is replaced entirely by our stratified BLR profile, built from the same Cloudy photoionization grid and Keplerian velocity field used above (Figure~\ref{fig:3073}).

Because the extended red and blue wings constrain the BLR shape parameters here, and because \(\alpha\) and \(f_{\rm vir}\) are, as before, degenerate, we calibrate \(f_{\rm vir}(\alpha)\) analytically rather than fitting it freely: for each trial \(\alpha\), \(f_{\rm vir}\) is set so that the model H\(\alpha\) velocity dispersion matches the observed value, \(\sigma_{\rm H\alpha}=1550\,{\rm km\,s^{-1}}\), exactly. We use the line dispersion \(\sigma\) rather than the FWHM for this calibration, since FWHM is not a well-defined, comparable quantity for a profile built from the superposition of several differently-weighted Gaussian components, whereas the flux-weighted variance \(\sigma^2\) remains additive and physically meaningful for the composite. With \(f_{\rm vir}\) calibrated in this way, \(\alpha\) is the only remaining H$\alpha$ shape parameter. The H$\beta$ and He~{\sc ii} line widths can then be predicted conditionally on the adopted value of \(\alpha\), without introducing additional line-specific parameters.

The resulting \(\chi^2(\alpha)\) curve is nearly flat, as found for the three LRDs, and the H$\alpha$ profile of GS-3073 alone therefore does not provide an independent constraint on \(\alpha\). We consequently adopt a Gaussian prior centered at \(\alpha=1.75\), with a dispersion of 0.25 corresponding to one grid step, motivated by the solutions obtained for GN-68797, GS-13971, and GN-9771. Including this prior yields an adopted solution with \(\alpha=1.75\) and \(f_{\rm vir}=1.1\). Thus, the value of \(\alpha\) is not independently measured from GS-3073; rather, the H$\alpha$ profile demonstrates that the radial stratification inferred from the three LRDs is also consistent with this unabsorbed LBD.

For this prior-informed solution, calibrated to reproduce the observed H$\alpha$ line dispersion of \(\sigma_{\rm H\alpha}=1550\,{\rm km\,s^{-1}}\), the model predicts \(\sigma_{\rm HeII}=2600\,{\rm km\,s^{-1}}\), corresponding to \(\sigma_{\rm HeII}/\sigma_{\rm H\alpha}=1.7\). The predicted H$\beta$ dispersion is \(1880\,{\rm km\,s^{-1}}\), or \(\sigma_{\rm H\beta}/\sigma_{\rm H\alpha}=1.2\). These are conditional predictions of the H$\alpha$-calibrated model and require no additional line-specific parameters. The model therefore predicts the observed sense of ionization stratification, with higher-ionization emission weighted toward smaller radii and larger characteristic velocities. Quantitatively, however, the predicted He~{\sc ii}-to-H$\alpha$ width ratio remains below the value of \(\sim2.1\pm 0.2\) implied by the observed dispersions. This shortfall persists across the full \(\alpha\) scan: no value of \(\alpha\) raises \(\sigma_{\rm HeII}/\sigma_{\rm H\alpha}\) above \(\sim\)1.7--1.8, because the Cloudy He~{\sc ii} emissivity turns over and declines at the highest ionization parameters sampled, while H\(\alpha\) emissivity continues to rise. This caps how much additional weight the model can place on the innermost, highest-velocity shells specifically for He~{\sc ii}, independent of the adopted radial slope.

\begin{figure*}[!htb]
\centering
\includegraphics[width=\hsize,trim=0 0 0 0,clip]{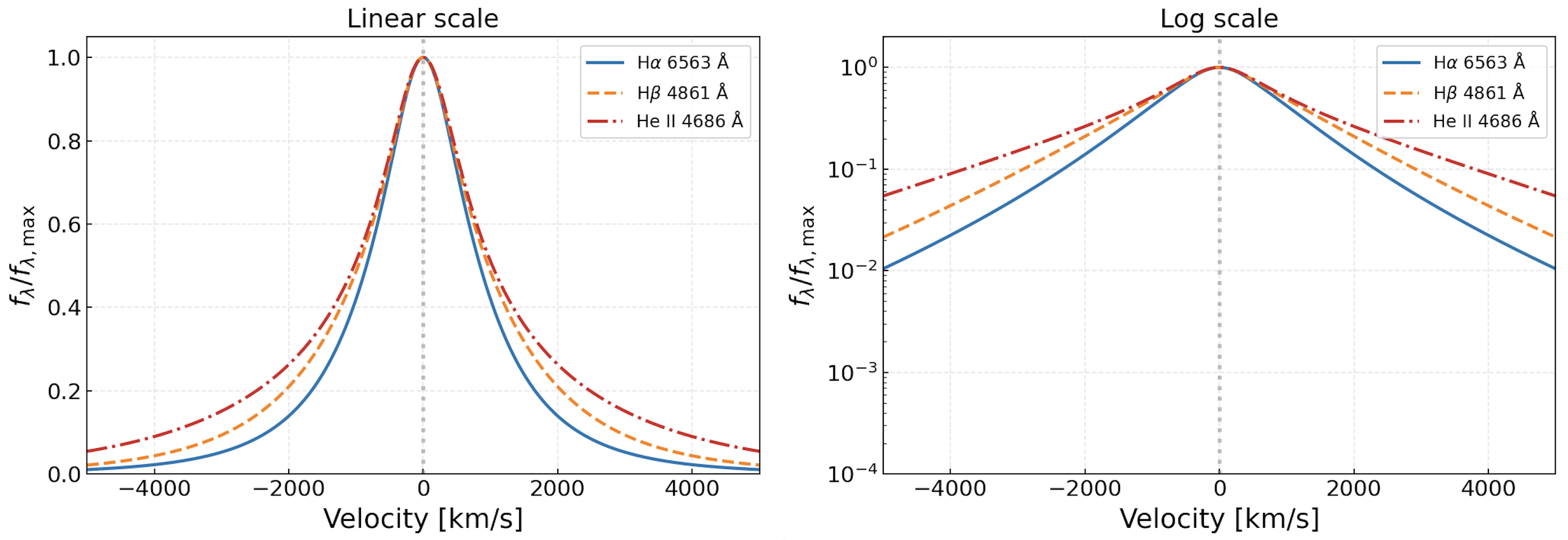}
\caption{
Predicted broad-line profiles for H\(\alpha\), H\(\beta\), and He~{\sc ii}~\(\lambda4686\) at the adopted stratified-BLR solution (\(\alpha=1.75\), \(f_{\rm vir}=1.1\)) of GS-3073. Left: linear scale; right: logarithmic scale, emphasizing the extended wings. All three profiles are computed from the same physical model and geometry constrained by the H\(\alpha\) wing fit alone -- 
conditional on the adopted \(\alpha\) and the H$\alpha$ calibration, the relative widths and wing strengths of H$\beta$ and He~{\sc ii} are predicted without additional line-specific parameters, with progressively broader, more extended profiles reflecting the decreasing Cloudy emissivity-weighted characteristic radius with increasing ionization potential.
}
\label{fig:predict}
\end{figure*}

Figure~\ref{fig:predict} shows the H$\alpha$, H$\beta$, and He~{\sc ii} profiles predicted by the prior-informed, H$\alpha$-calibrated solution. The relative H$\beta$ and He~{\sc ii} widths require no additional line-specific parameters. This is qualitatively consistent with reverberation-mapping studies, which generally find H\(\alpha\) lags comparable to or somewhat longer than H\(\beta\) lags -- i.e., H\(\alpha\) emission weighted toward slightly larger BLR radii than H\(\beta\) -- commonly attributed to differential optical depth through the Balmer series \citep[e.g.,][]{Grier2017}. A similar excess width of H\(\beta\) relative to H\(\alpha\), at marginal (\(\sim1.3\sigma\)) significance, has also been reported for the LRD GN-9771 (FWHM\(_{\rm H\beta}=2180\pm490\) vs FWHM\(_{\rm H\alpha}=1549\pm5\,{\rm km\,s^{-1}}\)), one of our own comparison sources (Section~\ref{sec:lrds}) \citep{Torralba2026b}. That work notes this result sits in tension with a pure electron-scattering origin for the Balmer wings, which predicts equal FWHM for lines scattered within the same layer, but does not propose an alternative explanation for the discrepancy. The stratified-BLR framework adopted here offers one: a broader H\(\beta\) emerges as a natural, parameter-free consequence of H\(\beta\)'s Cloudy emissivity being weighted toward marginally smaller, faster-moving radii than H\(\alpha\)'s.

\begin{table*}
\caption{Predicted broad, narrow, and total H$\alpha$ equivalent widths as a function of observer inclination for the three LRDs, adopting the best-fit radial slope $\alpha=1.75$, a BLR covering factor $C_{\rm BLR}=0.1$, and an angular thickness $\sigma_{\rm BLR}=0.2$. The narrow-to-broad H$\alpha$ flux ratios are fixed to the values inferred from the line-profile decompositions. Boldface highlights representative high-inclination solutions whose total EWs are closest to the observed values. Because the same $C_{\rm BLR}$, $\sigma_{\rm BLR}$, $\alpha$, and fiducial central engine are adopted for all three sources, the predicted broad H$\alpha$ EW depends only on observer inclination and is therefore identical across the three columns; the source-to-source differences in total EW arise solely from the different adopted narrow-to-broad flux ratios.
}
\label{tab:ew_theta}
\centering
\begin{tabular}{cccc|ccc|ccc}
\hline\hline
& \multicolumn{3}{c|}{GN-68797} & \multicolumn{3}{c|}{GS-13971} & \multicolumn{3}{c}{GN-9771} \\
$i_{\rm obs}$ & EW$_{\rm BLR}$ & EW$_{\rm NLR}$ & EW$_{\rm tot}$ & EW$_{\rm BLR}$ & EW$_{\rm NLR}$ & EW$_{\rm tot}$ & EW$_{\rm BLR}$ & EW$_{\rm NLR}$ & EW$_{\rm tot}$ \\
 & (\AA) & (\AA) & (\AA) & (\AA) & (\AA) & (\AA) & (\AA) & (\AA) & (\AA) \\
\hline
30$^\circ$ & 153 & 41  & 194  & 153 & 35  & 188  & 153 & 5  & 158  \\
50$^\circ$ & 220 & 59  & 279  & 220 & 51  & 271  & 220 & 7  & 227  \\
70$^\circ$ & 436 & 117 & 553  & 436 & 100 & 537  & 436 & 15 & 451  \\
75$^\circ$ & 577 & 154 & 732  & 577 & 133 & 710  & 577 & 19 & 597  \\
80$^\circ$ & \textbf{829} & \textbf{221} & \textbf{1051} & \textbf{829} & \textbf{190} & \textbf{1020} & 829 & 28 & 857 \\
85$^\circ$ & 1379 & 368 & 1747 & 1379 & 317 & 1695 & \textbf{1379} & \textbf{46} & \textbf{1425} \\
\hline
Observed & \multicolumn{3}{c|}{1306} & \multicolumn{3}{c|}{894} & \multicolumn{3}{c}{1713} \\
\hline
\end{tabular}
\end{table*}

\section{Discussion and summary}
\label{sec:summary}
So far we have focused on fitting the shape of the broad H$\alpha$ profile. If the BLR is flattened and its kinematics are dominated by ordered in-plane motion, then the large line widths inferred here are naturally suggestive of high observer inclinations, since the projected velocity field increases approximately as $\sin i_{\rm obs}$ \citep[e.g.,][]{Peterson2004,Pancoast2014,Netzer2020}. It is then natural to ask whether the same stratified-BLR framework, supplemented by representative choices for the BLR covering factor and angular thickness, can also account for the large observed H$\alpha$ EWs of the three LRDs. To this end, we evaluate the inclination dependence of the broad-line EW within the same model adopted for the profile fitting.

In our model, the BLR is described by an equatorially concentrated cloud distribution with covering factor $C_{\rm BLR}=0.1$ and angular thickness $\sigma_{\rm BLR}=0.2$. A global covering factor of $\simeq 0.1$ is consistent with standard estimates for low-$z$ Type~1 AGNs \citep[e.g.,][see also \citealt{MadauMaiolino2026}]{Peterson2006,Pandey2023}. This angular thickness implies a geometrically flattened BLR, in which the probability of intercepting clouds increases strongly toward the equatorial plane. The broad-line profile is produced by a radially stratified distribution of virialized clouds, while the observed EW depends on the inclination-dependent continuum against which the line is seen. We compute ${\rm EW}(i_{\rm obs})$ using the formalism of \citet{MadauMaiolino2026}, in which the broad-line and nebular continuum emission are treated as isotropic, while the direct and transmitted optical continuum increases toward lower observer inclinations. Because the direct continuum is progressively suppressed toward high inclinations, the model naturally predicts larger H$\alpha$ EWs for more edge-on sightlines \citep{Madau2026,MadauMaiolino2026}.

This trend is borne out by the calculations. Table~\ref{tab:ew_theta} shows the predicted broad, narrow, and total H$\alpha$ equivalent widths for a representative set of observer inclinations. For all three LRDs, the total EW increases steeply toward equatorial sightlines, reaching values comparable to those observed only for $i_{\rm obs}\gtrsim 80^\circ$. This provides an independent consistency check on the high-inclination interpretation suggested by the profile fitting and by the red continua of these sources. For the best-fit value $\alpha=1.75$, the broad H$\alpha$ EW rises from only $\sim 150\,$\AA\ for nearly polar sightlines to $\sim 580\,$\AA\ at $75^\circ$ and $\sim 1380\,$\AA\ at $85^\circ$. The observed total EWs of the three LRDs therefore fall naturally within the range expected for highly inclined observers. At fixed angular thickness, the predicted EW increases with BLR covering factor, so somewhat lower inclinations could in principle reproduce the observed values if $C_{\rm BLR}$ were larger than the representative value adopted here. Nevertheless, for a modest covering factor and a strongly equatorial angular distribution, the observed H$\alpha$ EWs are naturally matched only at high inclinations, consistent with the interpretation suggested independently by the red continua and absorption features.

The profile fitting leads to a similarly coherent picture. All three LRDs are well reproduced with the same radial slope, $\alpha=1.75$, indicating that the cloud distribution is concentrated near the dust sublimation radius, while the inner BLR still supplies the highest-velocity tails through the virial scaling $v_{\rm vir}\propto r^{-1/2}$. Although $\alpha$ is in principle partially degenerate with the adopted fiducial black-hole mass,\footnote{For a larger adopted $\Mh$, the virial velocities increase at all radii, which could be partially compensated by a larger $\alpha$ that shifts more weight toward the lower-velocity outer BLR; conversely, a smaller $\Mh$ would favor a flatter radial distribution.} in practice the three LRDs all converge to the same best-fit value, $\alpha\simeq 1.75$, whereas the source-to-source differences are captured mainly by $f_{\rm vir}$. This suggests that the preferred radial stratification is a stable feature of the fits.
The fitted values of the effective virial parameter provide an additional consistency check. For GN-68797 and GN-9771 we obtain $f_{\rm vir}=1.2$ and $f_{\rm vir}=1.3$, respectively, while for GS-13971 we find $f_{\rm vir}\simeq 0.94$. Thus, in one source the fiducial virial scaling is already sufficient, whereas in the other two a modest upward correction is required, plausibly reflecting a combination of geometry, kinematics, and the fact that the true black-hole mass may exceed the reference value adopted in the model.

The LBD GS-3073 extends this picture to a physically distinct regime. Unlike the three absorbed LRDs, GS-3073 shows a blue optical/UV continuum and no detected Balmer absorption, yet its broad H$\alpha$ wings are equally well described by the same stratified-BLR framework, and the adopted, prior-informed radial slope -- based on the \(\alpha=1.75\) solutions for GN-68797, GS-13971, and GN-9771 (Section~\ref{sec:gs3073}) -- is fully consistent with the value preferred by the three LRDs in isolation. Because $\alpha$ and $f_{\rm vir}$ are strongly degenerate -- including within GS-3073's own $\chi^2(\alpha)$ curve -- its H$\alpha$ profile alone cannot independently confirm $\alpha=1.75$; the value is adopted via the external prior, not derived from GS-3073 in isolation. What its data does show is that $\alpha=1.75$ is not excluded: it falls within the flat, near-degenerate region of the $\chi^2(\alpha)$ curve ($\chi^2/{\rm dof}\simeq 4$), statistically indistinguishable from the formal minimum. That a single slope is at least consistent with four sources spanning both the absorbed-LRD and unabsorbed-LBD classes is compatible with the stratification being a property of the BLR itself, though given the degeneracy this is a consistency argument, not an independent confirmation.

GS-3073 additionally offers a test unavailable for the three LRDs, since it has a detected broad He~{\sc ii}~$\lambda4686$ line: the same $\alpha=1.75$, $f_{\rm vir}=1.1$ solution predicts $\sigma_{\rm HeII}/\sigma_{\rm H\alpha}=1.7$ and $\sigma_{\rm H\beta}/\sigma_{\rm H\alpha}=1.2$, both in the sense expected from ionization stratification -- higher-ionization, more centrally-weighted lines being kinematically broader -- though the He~{\sc ii} ratio falls slightly short of the value ($\sim2.1\pm 0.2$) implied if the observed dispersions are taken at face value (Section~\ref{sec:gs3073}). We regard this as a genuine, falsifiable limitation of the single-population, uniform-density radial model rather than a failure of the overall picture, since the same architecture reproduces the H$\alpha$ wings, the common radial slope, and the correct sense of the He~{\sc ii} and H$\beta$ stratification without additional line-specific parameters. GS-3073's blue continuum and lack of Balmer absorption are also naturally accommodated within the same equatorial-wind geometry invoked below for the three LRDs: if the absorbing and reddening material is concentrated toward the equatorial plane, a source viewed at lower inclination -- away from the densest wind streamlines -- would be expected to show both a bluer continuum and weaker or absent Balmer absorption, exactly as observed for GS-3073, without requiring a different underlying BLR or wind structure.

Our results therefore support the following picture. A BLR with moderate covering factor, strong equatorial concentration, and radial stratification can simultaneously account for three otherwise puzzling features of LRDs: very large Balmer equivalent widths, broad exponential wings, and absorption-distorted cores. Within this framework, the large H$\alpha$ EWs do not require unusually large covering factors, but follow naturally from super-Eddington SEDs that provide a higher ionizing-to-optical photon budget than standard quasar composites, thereby boosting the intrinsic line-to-continuum ratio. Likewise, the broad wings need not be attributed primarily to Thomson scattering, but emerge from the radial superposition of virialized clouds. This does not exclude a minor contribution from electron scattering -- which may well operate in the dense inner regions of these systems -- but rather shows that it is not required as the dominant mechanism shaping the wings.

The blueshifted absorption troughs detected in all three LRDs are consistent with a P-Cygni interpretation, in which the absorbing gas has a net outflow component along the observer's line of sight. In super-Eddington accretion flows, radiatively driven material launched from the disc surface \citep[e.g.,][]{Murray1995,Proga2004} could naturally produce such blueshifted absorption. The absorber must lie at radii comparable to or larger than the BLR, since it imprints on the broad H$\alpha$ profile rather than on the continuum alone. Because the densest wind streamlines are concentrated near the equatorial plane, they are preferentially intercepted by high-inclination LRD sightlines, whereas lower-inclination LBD sightlines -- including GS-3073, as noted above -- cross a much smaller absorbing column, naturally accounting for the much weaker incidence of Balmer absorption in LBDs.

The fraction of little dots with genuinely exponential wings also merits comment in light of our results. \citet{Scholtz2026} find no statistical preference for an exponential profile in $\sim$60\% of their 32-object sample, with Lorentzian or multi-Gaussian models providing equally good or better fits, and note that LRDs specifically favor Lorentzian profiles more often than LBDs do. This is naturally accommodated within our stratified-BLR picture: a superposition of virialized clouds spanning a range of radii generically produces extended, non-Gaussian wings, but the precise functional form -- exponential, Lorentzian, or some intermediate shape -- depends on the detailed radial weighting of cloud number and emissivity, rather than being a fixed prediction of the model. The absence of a strictly exponential wing in a given source therefore does not argue against BLR stratification as the broadening mechanism; it simply reflects a different realization of the same underlying radial cloud distribution.

Although our stratified-BLR model does not by itself predict a one-to-one relation between Balmer-break strength and exponential-wing prominence \citep{Matthee2026}, a qualitative connection is natural: if stronger Balmer breaks trace a larger column of dense circumnuclear gas, they may also correspond to a larger effective BLR covering factor and/or a larger mass of BLR clouds along equatorial sightlines. The resulting increase in the number of virialized kinematic components contributing to the integrated profile would enhance the prominence of the nearly exponential wings. This qualitative connection should, however, be viewed in the context of the current observational uncertainty, since the relation between stronger Balmer breaks and more prominent exponential wings reported by \citet{Matthee2026} was not confirmed by \citet{Scholtz2026}. The three LRDs studied here, together with the LBD GS-3073, are consistent with a picture in which stratified BLRs viewed at high inclination can account for both the broad exponential wings and the large Balmer equivalent widths, while lower-inclination sightlines through the same underlying structure naturally account for unabsorbed, blue-continuum sources like GS-3073 -- providing a simple and physically plausible explanation for at least part of the LRD phenomenon.

\label{lastpage}
\bibliographystyle{aa}
\bibliography{paper}

@ARTICLE{Kollatschny2013,
       author = {{Kollatschny}, W. and {Zetzl}, M.},
        title = "{The shape of broad-line profiles in active galactic nuclei}",
      journal = {\aap},
         year = 2013,
        month = jan,
       volume = {549},
          eid = {A100},
        pages = {A100},
          doi = {10.1051/0004-6361/201219411},
archivePrefix = {arXiv},
       eprint = {1211.3065},
 primaryClass = {astro-ph.CO},
       adsurl = {https://ui.adsabs.harvard.edu/abs/2013A&A...549A.100K}
}

@ARTICLE{Scholtz2026,
       author = {{Scholtz}, J. and {D'Eugenio}, F. and {Maiolino}, R. and {Brazzini}, M. and {{\"U}bler}, H. and {Ji}, X. and {Perna}, M. and {Sun}, F. and {Brocchi}, G. and {Carniani}, S. and {Cresci}, G. and {Ivey}, L.~R. and {Juod{\v{z}}balis}, I. and {Marconi}, A. and {Mazzolari}, G. and {Risaliti}, G. and {Trefoloni}, B.},
        title = "{Little Red and Blue Dots: simply stratified Broad Line Regions}",
      journal = {arXiv e-prints},
         year = 2026,
        month = mar,
          eid = {arXiv:2603.22277},
        pages = {arXiv:2603.22277},
          doi = {10.48550/arXiv.2603.22277},
archivePrefix = {arXiv},
       eprint = {2603.22277},
 primaryClass = {astro-ph.GA},
       adsurl = {https://ui.adsabs.harvard.edu/abs/2026arXiv260322277S}
}

@ARTICLE{Matthee2026,
       author = {{Matthee}, Jorryt and {Torralba}, Alberto and {Pezzulli}, Gabriele and {Naidu}, Rohan P. and {Chisholm}, John and {Mascia}, Sara and {Greene}, Jenny E. and {Ishikawa}, Yuzo and {Gronke}, Max and {Wuyts}, Stijn and {Bordoloi}, Rongmon and {Brammer}, Gabriel and {Chang}, Seok-Jun and {Eilers}, Anna-Christina and {de Graaff}, Anna and {Hviding}, Raphael E. and {Iani}, Edoardo and {Illingworth}, Garth and {Kashino}, Daichi and {Labbe}, Ivo and {Ma}, Yilun and {Maseda}, Michael V. and {Meyer}, Romain and {Nelson}, Erica and {Oesch}, Pascal and {Xiao}, Mengyuan},
        title = "{The Engine and its Flows: Little Red Dot spectra are shaped by the column densities of their gas envelopes}",
      journal = {arXiv e-prints},
         year = 2026,
        month = mar,
          eid = {arXiv:2603.17667},
        pages = {arXiv:2603.17667},
          doi = {10.48550/arXiv.2603.17667},
archivePrefix = {arXiv},
       eprint = {2603.17667},
 primaryClass = {astro-ph.GA},
       adsurl = {https://ui.adsabs.harvard.edu/abs/2026arXiv260317667M}
}

@ARTICLE{MadauMaiolino2026,
       author = {{Madau}, Piero and {Maiolino}, Roberto},
        title = "{Little Red Dots as Obscured Little Blue Dots: A Super-Eddington Unification Model}",
      journal = {arXiv e-prints},
         year = 2026,
        month = feb,
          eid = {arXiv:2602.22386},
        pages = {arXiv:2602.22386},
          doi = {10.48550/arXiv.2602.22386},
archivePrefix = {arXiv},
       eprint = {2602.22386},
 primaryClass = {astro-ph.GA},
       adsurl = {https://ui.adsabs.harvard.edu/abs/2026arXiv260222386M}
}

@ARTICLE{Pacucci2026,
       author = {{Pacucci}, Fabio and {Ferrara}, Andrea and {Kocevski}, Dale D.},
        title = "{The Little Red Dots Are Direct Collapse Black Holes}",
      journal = {arXiv e-prints},
         year = 2026,
        month = jan,
          eid = {arXiv:2601.14368},
        pages = {arXiv:2601.14368},
          doi = {10.48550/arXiv.2601.14368},
archivePrefix = {arXiv},
       eprint = {2601.14368},
 primaryClass = {astro-ph.GA},
       adsurl = {https://ui.adsabs.harvard.edu/abs/2026arXiv260114368P}
}

@ARTICLE{NetzerLaor1993,
       author = {{Netzer}, Hagai and {Laor}, Ari},
        title = "{Dust in the Narrow-Line Region of Active Galactic Nuclei}",
      journal = {\apjl},
         year = 1993,
        month = feb,
       volume = {404},
        pages = {L51},
          doi = {10.1086/186741},
       adsurl = {https://ui.adsabs.harvard.edu/abs/1993ApJ...404L..51N}
}

@ARTICLE{Juod2026,
       author = {{Juod{\v{z}}balis}, Ignas and {Maiolino}, Roberto and {Baker}, William M. and {Lake}, Emma Curtis and {Scholtz}, Jan and {D'Eugenio}, Francesco and {Trefoloni}, Bartolomeo and {Isobe}, Yuki and {Tacchella}, Sandro and {Bunker}, Andrew J. and {Carniani}, Stefano and {Charlot}, St{\'e}phane and {Jones}, Gareth C. and {Parlanti}, Eleonora and {Perna}, Michele and {Rinaldi}, Pierluigi and {Robertson}, Brant and {{\"U}bler}, Hannah and {Venturi}, Giacomo and {Willott}, Chris},
        title = "{JADES: comprehensive census of broad-line AGN from reionization to cosmic noon revealed by JWST}",
      journal = {\mnras},
         year = 2026,
        month = mar,
       volume = {546},
       number = {3},
          eid = {stag086},
        pages = {stag086},
          doi = {10.1093/mnras/stag086},
archivePrefix = {arXiv},
       eprint = {2504.03551},
 primaryClass = {astro-ph.GA},
       adsurl = {https://ui.adsabs.harvard.edu/abs/2026MNRAS.546ag086J}
}

@ARTICLE{Kocevski2025,
       author = {{Kocevski}, Dale D. and {Finkelstein}, Steven L. and {Barro}, Guillermo and {Taylor}, Anthony J. and {Calabr{\`o}}, Antonello and {Laloux}, Brivael and {Buchner}, Johannes and {Trump}, Jonathan R. and {Leung}, Gene C.~K. and {Yang}, Guang and {Dickinson}, Mark and {P{\'e}rez-Gonz{\'a}lez}, Pablo G. and {Pacucci}, Fabio and {Inayoshi}, Kohei and {Somerville}, Rachel S. and {McGrath}, Elizabeth J. and {Akins}, Hollis B. and {Bagley}, Micaela B. and {Bowler}, Rebecca A.~A. and {Bisigello}, Laura and {Carnall}, Adam and {Casey}, Caitlin M. and {Cheng}, Yingjie and {Cleri}, Nikko J. and {Costantin}, Luca and {Cullen}, Fergus and {Davis}, Kelcey and {Donnan}, Callum T. and {Dunlop}, James S. and {Ellis}, Richard S. and {Ferguson}, Henry C. and {Fujimoto}, Seiji and {Fontana}, Adriano and {Giavalisco}, Mauro and {Grazian}, Andrea and {Grogin}, Norman A. and {Hathi}, Nimish P. and {Hirschmann}, Michaela and {Huertas-Company}, Marc and {Holwerda}, Benne W. and {Illingworth}, Garth and {Juneau}, St{\'e}phanie and {Kartaltepe}, Jeyhan S. and {Koekemoer}, Anton M. and {Li}, Wenxiu and {Lucas}, Ray A. and {Magee}, Dan and {Mason}, Charlotte and {McLeod}, Derek J. and {McLure}, Ross J. and {Napolitano}, Lorenzo and {Papovich}, Casey and {Pirzkal}, Nor and {Rodighiero}, Giulia and {Santini}, Paola and {Wilkins}, Stephen M. and {Yung}, L.~Y. Aaron},
        title = "{The Rise of Faint, Red Active Galactic Nuclei at z > 4: A Sample of Little Red Dots in the JWST Extragalactic Legacy Fields}",
      journal = {\apj},
         year = 2025,
        month = jun,
       volume = {986},
       number = {2},
          eid = {126},
        pages = {126},
          doi = {10.3847/1538-4357/adbc7d},
archivePrefix = {arXiv},
       eprint = {2404.03576},
 primaryClass = {astro-ph.GA},
       adsurl = {https://ui.adsabs.harvard.edu/abs/2025ApJ...986..126K}
}

@ARTICLE{Rusakov2026,
       author = {{Rusakov}, V. and {Watson}, D. and {Nikopoulos}, G.~P. and {Brammer}, G. and {Gottumukkala}, R. and {Harvey}, T. and {Heintz}, K.~E. and {Damgaard}, R. and {Sim}, S.~A. and {Sneppen}, A. and {Vijayan}, A.~P. and {Adams}, N. and {Austin}, D. and {Conselice}, C.~J. and {Goolsby}, C.~M. and {Toft}, S. and {Witstok}, J.},
        title = "{Little red dots as young supermassive black holes in dense ionized cocoons}",
      journal = {\nat},
         year = 2026,
        month = jan,
       volume = {649},
       number = {8097},
        pages = {574-579},
          doi = {10.1038/s41586-025-09900-4},
       adsurl = {https://ui.adsabs.harvard.edu/abs/2026Natur.649..574R}
}

@ARTICLE{Pandey2023,
       author = {{Pandey}, Ashwani and {Czerny}, Bo{\.z}ena and {Panda}, Swayamtrupta and {Prince}, Raj and {Jaiswal}, Vikram Kumar and {Martinez-Aldama}, Mary Loli and {Zaja{\v{c}}ek}, Michal and {{\'S}niegowska}, Marzena},
        title = "{Broad-line region in active galactic nuclei: Dusty or dustless?}",
      journal = {\aap},
         year = 2023,
        month = dec,
       volume = {680},
          eid = {A102},
        pages = {A102},
          doi = {10.1051/0004-6361/202347819},
archivePrefix = {arXiv},
       eprint = {2310.05089},
 primaryClass = {astro-ph.GA},
       adsurl = {https://ui.adsabs.harvard.edu/abs/2023A&A...680A.102P}
}

@ARTICLE{Brazzini2026,
       author = {{Brazzini}, M. and {D'Eugenio}, F. and {Maiolino}, R. and {Lyu}, J. and {DeCoursey}, C. and {{\"U}bler}, H. and {Ji}, X. and {Juod{\v{z}}balis}, I. and {Scholtz}, J. and {Jones}, G.~C. and {Hainline}, K. and {Dalla Bont{\`a}}, E. and {{\'e}rez-Gonz{\'a}lez}, P.~G. P and {Geris}, S. and {Harshan}, A. and {Feruglio}, C. and {Bischetti}, M. and {Mazzolari}, G. and {Rieke}, G. and {Alberts}, S. and {Trefoloni}, B. and {Carniani}, S. and {Parlanti}, E. and {Marconi}, A. and {Risaliti}, G. and {Ramos Almeida}, C. and {Rinaldi}, P. and {Perna}, M. and {Zamora}, S. and {Lamperti}, I. and {Venturi}, G. and {Cresci}, G. and {Bunker}, Andrew J. and {Ivey}, L.~R.},
        title = "{The Little Blue and Red Dots Rosetta Stones: Non-Gaussian broad lines, hot dust, and X-ray weakness}",
      journal = {arXiv e-prints},
         year = 2026,
        month = jan,
          eid = {arXiv:2601.22214},
        pages = {arXiv:2601.22214},
          doi = {10.48550/arXiv.2601.22214},
archivePrefix = {arXiv},
       eprint = {2601.22214},
 primaryClass = {astro-ph.GA},
       adsurl = {https://ui.adsabs.harvard.edu/abs/2026arXiv260122214B}
}

@ARTICLE{Laor2006,
       author = {{Laor}, Ari},
        title = "{Evidence for Line Broadening by Electron Scattering in the Broad-Line Region of NGC 4395}",
      journal = {\apj},
         year = 2006,
        month = may,
       volume = {643},
       number = {1},
        pages = {112-119},
          doi = {10.1086/502798},
archivePrefix = {arXiv},
       eprint = {astro-ph/0601688},
 primaryClass = {astro-ph},
       adsurl = {https://ui.adsabs.harvard.edu/abs/2006ApJ...643..112L}
}

@ARTICLE{Delvecchio2025,
       author = {{Delvecchio}, I. and {Daddi}, E. and {Magnelli}, B. and {Elbaz}, D. and {Giavalisco}, M. and {Traina}, A. and {Lanzuisi}, G. and {Akins}, H.~B. and {Belli}, S. and {Casey}, C.~M. and {Gentile}, F. and {Gruppioni}, C. and {Pozzi}, F. and {Zamorani}, G.},
        title = "{Active galactic nuclei-heated dust revealed in ``little red dots''}",
      journal = {\aap},
         year = 2025,
        month = dec,
       volume = {704},
          eid = {A313},
        pages = {A313},
          doi = {10.1051/0004-6361/202557164},
archivePrefix = {arXiv},
       eprint = {2509.07100},
 primaryClass = {astro-ph.GA},
       adsurl = {https://ui.adsabs.harvard.edu/abs/2025A&A...704A.313D}
}

@ARTICLE{Akins2025a,
       author = {{Akins}, Hollis B. and {Casey}, Caitlin M. and {Lambrides}, Erini and {Allen}, Natalie and {Andika}, Irham T. and {Brinch}, Malte and {Champagne}, Jaclyn B. and {Cooper}, Olivia and {Ding}, Xuheng and {Drakos}, Nicole E. and {Faisst}, Andreas and {Finkelstein}, Steven L. and {Franco}, Maximilien and {Fujimoto}, Seiji and {Gentile}, Fabrizio and {Gillman}, Steven and {Gozaliasl}, Ghassem and {Harish}, Santosh and {Hayward}, Christopher C. and {Hirschmann}, Michaela and {Ilbert}, Olivier and {Kartaltepe}, Jeyhan S. and {Kocevski}, Dale D. and {Koekemoer}, Anton M. and {Kokorev}, Vasily and {Liu}, Daizhong and {Long}, Arianna S. and {McCracken}, Henry Joy and {McKinney}, Jed and {Onoue}, Masafusa and {Paquereau}, Louise and {Renzini}, Alvio and {Rhodes}, Jason and {Robertson}, Brant E. and {Shuntov}, Marko and {Silverman}, John D. and {Tanaka}, Takumi S. and {Toft}, Sune and {Trakhtenbrot}, Benny and {Valentino}, Francesco and {Zavala}, Jorge},
        title = "{COSMOS-Web: The Overabundance and Physical Nature of ``Little Red Dots''{\textemdash}Implications for Early Galaxy and SMBH Assembly}",
      journal = {\apj},
         year = 2025,
        month = sep,
       volume = {991},
       number = {1},
          eid = {37},
        pages = {37},
          doi = {10.3847/1538-4357/ade984},
archivePrefix = {arXiv},
       eprint = {2406.10341},
 primaryClass = {astro-ph.GA},
       adsurl = {https://ui.adsabs.harvard.edu/abs/2025ApJ...991...37A}
}

@ARTICLE{Greene2024,
       author = {{Greene}, Jenny E. and {Labbe}, Ivo and {Goulding}, Andy D. and {Furtak}, Lukas J. and {Chemerynska}, Iryna and {Kokorev}, Vasily and {Dayal}, Pratika and {Volonteri}, Marta and {Williams}, Christina C. and {Wang}, Bingjie and {Setton}, David J. and {Burgasser}, Adam J. and {Bezanson}, Rachel and {Atek}, Hakim and {Brammer}, Gabriel and {Cutler}, Sam E. and {Feldmann}, Robert and {Fujimoto}, Seiji and {Glazebrook}, Karl and {de Graaff}, Anna and {Khullar}, Gourav and {Leja}, Joel and {Marchesini}, Danilo and {Maseda}, Michael V. and {Matthee}, Jorryt and {Miller}, Tim B. and {Naidu}, Rohan P. and {Nanayakkara}, Themiya and {Oesch}, Pascal A. and {Pan}, Richard and {Papovich}, Casey and {Price}, Sedona H. and {van Dokkum}, Pieter and {Weaver}, John R. and {Whitaker}, Katherine E. and {Zitrin}, Adi},
        title = "{UNCOVER Spectroscopy Confirms the Surprising Ubiquity of Active Galactic Nuclei in Red Sources at z > 5}",
      journal = {\apj},
         year = 2024,
        month = mar,
       volume = {964},
       number = {1},
          eid = {39},
        pages = {39},
          doi = {10.3847/1538-4357/ad1e5f},
archivePrefix = {arXiv},
       eprint = {2309.05714},
 primaryClass = {astro-ph.GA},
       adsurl = {https://ui.adsabs.harvard.edu/abs/2024ApJ...964...39G}
}

@ARTICLE{Kido2025,
       author = {{Kido}, Daisaburo and {Ioka}, Kunihito and {Hotokezaka}, Kenta and {Inayoshi}, Kohei and {Irwin}, Christopher M.},
        title = "{Black hole envelopes in Little Red Dots}",
      journal = {\mnras},
         year = 2025,
        month = dec,
       volume = {544},
       number = {4},
        pages = {3407-3416},
          doi = {10.1093/mnras/staf1898},
archivePrefix = {arXiv},
       eprint = {2505.06965},
 primaryClass = {astro-ph.HE},
       adsurl = {https://ui.adsabs.harvard.edu/abs/2025MNRAS.544.3407K}
}

@ARTICLE{Naidu2025,
       author = {{Naidu}, Rohan P. and {Matthee}, Jorryt and {Katz}, Harley and {de Graaff}, Anna and {Oesch}, Pascal and {Smith}, Aaron and {Greene}, Jenny E. and {Brammer}, Gabriel and {Weibel}, Andrea and {Hviding}, Raphael and {Chisholm}, John and {Labb\textbackslash'e}, Ivo and {Simcoe}, Robert A. and {Witten}, Callum and {Atek}, Hakim and {Baggen}, Josephine F.~W. and {Belli}, Sirio and {Bezanson}, Rachel and {Boogaard}, Leindert A. and {Bose}, Sownak and {Covelo-Paz}, Alba and {Dayal}, Pratika and {Fudamoto}, Yoshinobu and {Furtak}, Lukas J. and {Giovinazzo}, Emma and {Goulding}, Andy and {Gronke}, Max and {Heintz}, Kasper E. and {Hirschmann}, Michaela and {Illingworth}, Garth and {Inoue}, Akio K. and {Johnson}, Benjamin D. and {Leja}, Joel and {Leonova}, Ecaterina and {McConachie}, Ian and {Maseda}, Michael V. and {Natarajan}, Priyamvada and {Nelson}, Erica and {Setton}, David J. and {Shivaei}, Irene and {Sobral}, David and {Stefanon}, Mauro and {Tacchella}, Sandro and {Toft}, Sune and {Torralba}, Alberto and {van Dokkum}, Pieter and {van der Wel}, Arjen and {Volonteri}, Marta and {Walter}, Fabian and {Wang}, Bingjie and {Watson}, Darach},
        title = "{A ``Black Hole Star'' Reveals the Remarkable Gas-Enshrouded Hearts of the Little Red Dots}",
      journal = {arXiv e-prints},
         year = 2025,
        month = mar,
          eid = {arXiv:2503.16596},
        pages = {arXiv:2503.16596},
          doi = {10.48550/arXiv.2503.16596},
archivePrefix = {arXiv},
       eprint = {2503.16596},
 primaryClass = {astro-ph.GA},
       adsurl = {https://ui.adsabs.harvard.edu/abs/2025arXiv250316596N}
}

@ARTICLE{Chatzikos2023,
       author = {{Chatzikos}, M. and {Bianchi}, S. and {Camilloni}, F. and {Chakraborty}, P. and {Gunasekera}, C.~M. and {Guzm{\'a}n}, F. and {Milby}, J.~S. and {Sarkar}, A. and {Shaw}, G. and {van Hoof}, P.~A.~M. and {Ferland}, G.~J.},
        title = "{The 2023 Release of Cloudy}",
      journal = {\rmxaa},
         year = 2023,
        month = oct,
       volume = {59},
        pages = {327-343},
          doi = {10.22201/ia.01851101p.2023.59.02.12},
archivePrefix = {arXiv},
       eprint = {2308.06396},
 primaryClass = {astro-ph.GA},
       adsurl = {https://ui.adsabs.harvard.edu/abs/2023RMxAA..59..327C}
}

@ARTICLE{Hainline2025,
       author = {{Hainline}, Kevin N. and {Maiolino}, Roberto and {Juod{\v{z}}balis}, Ignas and {Scholtz}, Jan and {{\"U}bler}, Hannah and {D'Eugenio}, Francesco and {Helton}, Jakob M. and {Sun}, Yang and {Sun}, Fengwu and {Robertson}, Brant and {Tacchella}, Sandro and {Bunker}, Andrew J. and {Carniani}, Stefano and {Charlot}, Stephane and {Curtis-Lake}, Emma and {Egami}, Eiichi and {Johnson}, Benjamin D. and {Lin}, Xiaojing and {Lyu}, Jianwei and {P{\'e}rez-Gonz{\'a}lez}, Pablo G. and {Rinaldi}, Pierluigi and {Silcock}, Maddie S. and {Venturi}, Giacomo and {Williams}, Christina C. and {Willmer}, Christopher N.~A. and {Willott}, Chris and {Zhang}, Junyu and {Zhu}, Yongda},
        title = "{An Investigation into the Selection and Colors of Little Red Dots and Active Galactic Nuclei}",
      journal = {\apj},
         year = 2025,
        month = feb,
       volume = {979},
       number = {2},
          eid = {138},
        pages = {138},
          doi = {10.3847/1538-4357/ad9920},
archivePrefix = {arXiv},
       eprint = {2410.00100},
 primaryClass = {astro-ph.GA},
       adsurl = {https://ui.adsabs.harvard.edu/abs/2025ApJ...979..138H}
}

@ARTICLE{Hviding2025,
       author = {{Hviding}, Raphael E. and {de Graaff}, Anna and {Miller}, Tim B. and {Setton}, David J. and {Greene}, Jenny E. and {Labb{\'e}}, Ivo and {Brammer}, Gabriel and {Bezanson}, Rachel and {Boogaard}, Leindert A. and {Cleri}, Nikko J. and {Leja}, Joel and {Maseda}, Michael V. and {McConachie}, Ian and {Matthee}, Jorryt and {Naidu}, Rohan P. and {Oesch}, Pascal A. and {Wang}, Bingjie and {Whitaker}, Katherine E. and {Williams}, Christina C.},
        title = "{RUBIES: A spectroscopic census of little red dots: All point sources with v-shaped continua have broad lines}",
      journal = {\aap},
         year = 2025,
        month = oct,
       volume = {702},
          eid = {A57},
        pages = {A57},
          doi = {10.1051/0004-6361/202555816},
archivePrefix = {arXiv},
       eprint = {2506.05459},
 primaryClass = {astro-ph.GA},
       adsurl = {https://ui.adsabs.harvard.edu/abs/2025A&A...702A..57H}
}

@ARTICLE{Tang2025,
       author = {{Tang}, Mengtao and {Stark}, Daniel P. and {Plat}, Ad{\`e}le and {Feltre}, Anna and {Katz}, Harley and {Senchyna}, Peter and {Mason}, Charlotte A. and {Whitler}, Lily and {Chen}, Zuyi and {Topping}, Michael W.},
        title = "{JWST/NIRSpec Observations of High-ionization Emission Lines in Galaxies at High Redshift}",
      journal = {\apj},
         year = 2025,
        month = oct,
       volume = {991},
       number = {2},
          eid = {217},
        pages = {217},
          doi = {10.3847/1538-4357/adfd57},
archivePrefix = {arXiv},
       eprint = {2505.06359},
 primaryClass = {astro-ph.GA},
       adsurl = {https://ui.adsabs.harvard.edu/abs/2025ApJ...991..217T}
}

@ARTICLE{Harikane2023AGN,
       author = {{Harikane}, Yuichi and {Zhang}, Yechi and {Nakajima}, Kimihiko and {Ouchi}, Masami and {Isobe}, Yuki and {Ono}, Yoshiaki and {Hatano}, Shun and {Xu}, Yi and {Umeda}, Hiroya},
        title = "{A JWST/NIRSpec First Census of Broad-line AGNs at z = 4-7: Detection of 10 Faint AGNs with M $_{BH}$ {}10$^{6}$-{}10$^{8}$ M $_{{\ensuremath{\odot}}}$ and Their Host Galaxy Properties}",
      journal = {\apj},
         year = 2023,
        month = dec,
       volume = {959},
       number = {1},
          eid = {39},
        pages = {39},
          doi = {10.3847/1538-4357/ad029e},
archivePrefix = {arXiv},
       eprint = {2303.11946},
 primaryClass = {astro-ph.GA},
       adsurl = {https://ui.adsabs.harvard.edu/abs/2023ApJ...959...39H}
}

@ARTICLE{Peterson1993,
       author = {{Peterson}, Bradley M.},
        title = "{Reverberation Mapping of Active Galactic Nuclei}",
      journal = {\pasp},
         year = 1993,
        month = mar,
       volume = {105},
        pages = {247},
          doi = {10.1086/133140},
       adsurl = {https://ui.adsabs.harvard.edu/abs/1993PASP..105..247P}
}

@ARTICLE{Pancoast2014,
       author = {{Pancoast}, Anna and {Brewer}, Brendon J. and {Treu}, Tommaso},
        title = "{Modelling reverberation mapping data - I. Improved geometric and dynamical models and comparison with cross-correlation results}",
      journal = {\mnras},
         year = 2014,
        month = dec,
       volume = {445},
       number = {3},
        pages = {3055-3072},
          doi = {10.1093/mnras/stu1809},
archivePrefix = {arXiv},
       eprint = {1407.2941},
 primaryClass = {astro-ph.GA},
       adsurl = {https://ui.adsabs.harvard.edu/abs/2014MNRAS.445.3055P}
}

@ARTICLE{Grier2017,
       author = {{Grier}, C.~J. and {Trump}, J.~R. and {Shen}, Yue and {Horne}, Keith and {Kinemuchi}, Karen and {McGreer}, Ian D. and {Starkey}, D.~A. and {Brandt}, W.~N. and {Hall}, P.~B. and {Kochanek}, C.~S. and {Chen}, Yuguang and {Denney}, K.~D. and {Greene}, Jenny E. and {Ho}, L.~C. and {Homayouni}, Y. and {I-Hsiu Li}, Jennifer and {Pei}, Liuyi and {Peterson}, B.~M. and {Petitjean}, P. and {Schneider}, D.~P. and {Sun}, Mouyuan and {AlSayyad}, Yusura and {Bizyaev}, Dmitry and {Brinkmann}, Jonathan and {Brownstein}, Joel R. and {Bundy}, Kevin and {Dawson}, K.~S. and {Eftekharzadeh}, Sarah and {Fernandez-Trincado}, J.~G. and {Gao}, Yang and {Hutchinson}, Timothy A. and {Jia}, Siyao and {Jiang}, Linhua and {Oravetz}, Daniel and {Pan}, Kaike and {Paris}, Isabelle and {Ponder}, Kara A. and {Peters}, Christina and {Rogerson}, Jesse and {Simmons}, Audrey and {Smith}, Robyn and {Wang}, Ran},
        title = "{The Sloan Digital Sky Survey Reverberation Mapping Project: H{\ensuremath{\alpha}} and H{\ensuremath{\beta}} Reverberation Measurements from First-year Spectroscopy and Photometry}",
      journal = {\apj},
         year = 2017,
        month = dec,
       volume = {851},
       number = {1},
          eid = {21},
        pages = {21},
          doi = {10.3847/1538-4357/aa98dc},
archivePrefix = {arXiv},
       eprint = {1711.03114},
 primaryClass = {astro-ph.GA},
       adsurl = {https://ui.adsabs.harvard.edu/abs/2017ApJ...851...21G}
}

@ARTICLE{Netzer2020,
       author = {{Netzer}, Hagai},
        title = "{Testing broad-line region models with reverberation mapping}",
      journal = {\mnras},
         year = 2020,
        month = may,
       volume = {494},
       number = {2},
        pages = {1611-1621},
          doi = {10.1093/mnras/staa767},
archivePrefix = {arXiv},
       eprint = {2003.07660},
 primaryClass = {astro-ph.GA},
       adsurl = {https://ui.adsabs.harvard.edu/abs/2020MNRAS.494.1611N}
}

@ARTICLE{Goad2012,
       author = {{Goad}, M.~R. and {Korista}, K.~T. and {Ruff}, A.~J.},
        title = "{The broad emission-line region: the confluence of the outer accretion disc with the inner edge of the dusty torus}",
      journal = {\mnras},
         year = 2012,
        month = nov,
       volume = {426},
       number = {4},
        pages = {3086-3111},
          doi = {10.1111/j.1365-2966.2012.21808.x},
archivePrefix = {arXiv},
       eprint = {1207.6339},
 primaryClass = {astro-ph.CO},
       adsurl = {https://ui.adsabs.harvard.edu/abs/2012MNRAS.426.3086G}
}

@ARTICLE{Barro2026,
       author = {{Barro}, Guillermo and {P{\'e}rez-Gonz{\'a}lez}, Pablo G. and {Kocevski}, Dale D. and {McGrath}, Elizabeth J. and {Leung}, Gene C.~K. and {Cullen}, Fergus and {Dunlop}, James S. and {Ellis}, Richard S. and {Finkelstein}, Steven L. and {Grogin}, Norman A. and {Illingworth}, Garth and {Kartaltepe}, Jeyhan S. and {Koekemoer}, Anton M. and {Lucas}, Ray A. and {McLure}, Ross J. and {Yang}, Guang},
        title = "{A Comprehensive Photometric Selection of ``Little Red Dots'' in MIRI Fields: An Infrared-Bright Little Red Dot at z = 3.1386 with Warm Dust Emission}",
      journal = {\apj},
         year = 2026,
        month = jan,
       volume = {997},
       number = {1},
          eid = {48},
        pages = {48},
          doi = {10.3847/1538-4357/ae0704},
archivePrefix = {arXiv},
       eprint = {2412.01887},
 primaryClass = {astro-ph.GA},
       adsurl = {https://ui.adsabs.harvard.edu/abs/2026ApJ...997...48B}
}

@ARTICLE{Peterson2004,
       author = {{Peterson}, B.~M. and {Ferrarese}, L. and {Gilbert}, K.~M. and {Kaspi}, S. and {Malkan}, M.~A. and {Maoz}, D. and {Merritt}, D. and {Netzer}, H. and {Onken}, C.~A. and {Pogge}, R.~W. and {Vestergaard}, M. and {Wandel}, A.},
        title = "{Central Masses and Broad-Line Region Sizes of Active Galactic Nuclei. II. A Homogeneous Analysis of a Large Reverberation-Mapping Database}",
      journal = {\apj},
         year = 2004,
        month = oct,
       volume = {613},
       number = {2},
        pages = {682-699},
          doi = {10.1086/423269},
archivePrefix = {arXiv},
       eprint = {astro-ph/0407299},
 primaryClass = {astro-ph},
       adsurl = {https://ui.adsabs.harvard.edu/abs/2004ApJ...613..682P}
}

@ARTICLE{Barvainis1987,
  author       = {Barvainis, R.},
  title        = {Hot Dust and the Near-Infrared Bump in the Continuum of Quasars},
  journal      = {Astrophysical Journal},
  year         = {1987},
  volume       = {320},
  pages        = {537},
  doi          = {10.1086/165544},
  adsurl       = {https://ui.adsabs.harvard.edu/abs/1987ApJ...320..537B}
}

@ARTICLE{Lambrides2024,
       author = {{Lambrides}, Erini and {Garofali}, Kristen and {Larson}, Rebecca and {Ptak}, Andrew and {Chiaberge}, Marco and {Long}, Arianna S. and {Hutchison}, Taylor A. and {Norman}, Colin and {McKinney}, Jed and {Akins}, Hollis B. and {Berg}, Danielle A. and {Chisholm}, John and {Civano}, Francesca and {Cloonan}, Aidan P. and {Endsley}, Ryan and {Faisst}, Andreas L. and {Gilli}, Roberto and {Gillman}, Steven and {Hirschmann}, Michaela and {Kartaltepe}, Jeyhan S. and {Kocevski}, Dale D. and {Kokorev}, Vasily and {Pacucci}, Fabio and {Richardson}, Chris T. and {Stiavelli}, Massimo and {Whalen}, Kelly E.},
        title = "{The Case for Super-Eddington Accretion: Connecting Weak X-ray and UV Line Emission in JWST Broad-Line AGN During the First Gyr of Cosmic Time}",
      journal = {arXiv e-prints},
         year = 2024,
        month = sep,
          eid = {arXiv:2409.13047},
        pages = {arXiv:2409.13047},
          doi = {10.48550/arXiv.2409.13047},
archivePrefix = {arXiv},
       eprint = {2409.13047},
 primaryClass = {astro-ph.HE},
       adsurl = {https://ui.adsabs.harvard.edu/abs/2024arXiv240913047L}
}

@ARTICLE{Lupi2024b,
       author = {{Lupi}, Alessandro and {Trinca}, Alessandro and {Volonteri}, Marta and {Dotti}, Massimo and {Mazzucchelli}, Chiara},
        title = "{Size matters: are we witnessing super-Eddington accretion in high-redshift black holes from JWST?}",
      journal = {\aap},
         year = 2024,
        month = sep,
       volume = {689},
          eid = {A128},
        pages = {A128},
          doi = {10.1051/0004-6361/202451249},
archivePrefix = {arXiv},
       eprint = {2406.17847},
 primaryClass = {astro-ph.HE},
       adsurl = {https://ui.adsabs.harvard.edu/abs/2024A&A...689A.128L}
}

@ARTICLE{Madau2024,
       author = {{Madau}, Piero and {Haardt}, Francesco},
        title = "{X-Ray Weak Active Galactic Nuclei from Super-Eddington Accretion onto Infant Black Holes}",
      journal = {\apjl},
         year = 2024,
        month = dec,
       volume = {976},
       number = {2},
          eid = {L24},
        pages = {L24},
          doi = {10.3847/2041-8213/ad90e1},
       adsurl = {https://ui.adsabs.harvard.edu/abs/2024ApJ...976L..24M}
}

@ARTICLE{Fausnaugh2017,
       author = {{Fausnaugh}, M.~M. and {Grier}, C.~J. and {Bentz}, M.~C. and {Denney}, K.~D. and {De Rosa}, G. and {Peterson}, B.~M. and {Kochanek}, C.~S. and {Pogge}, R.~W. and {Adams}, S.~M. and {Barth}, A.~J. and {Beatty}, Thomas G. and {Bhattacharjee}, A. and {Borman}, G.~A. and {Boroson}, T.~A. and {Bottorff}, M.~C. and {Brown}, Jacob E. and {Brown}, Jonathan S. and {Brotherton}, M.~S. and {Coker}, C.~T. and {Crawford}, S.~M. and {Croxall}, K.~V. and {Eftekharzadeh}, Sarah and {Eracleous}, Michael and {Joner}, M.~D. and {Henderson}, C.~B. and {Holoien}, T.~W.-S. and {Horne}, Keith and {Hutchison}, T. and {Kaspi}, Shai and {Kim}, S. and {King}, Anthea L. and {Li}, Miao and {Lochhaas}, Cassandra and {Ma}, Zhiyuan and {MacInnis}, F. and {Manne-Nicholas}, E.~R. and {Mason}, M. and {Montuori}, Carmen and {Mosquera}, Ana and {Mudd}, Dale and {Musso}, R. and {Nazarov}, S.~V. and {Nguyen}, M.~L. and {Okhmat}, D.~N. and {Onken}, Christopher A. and {Ou-Yang}, B. and {Pancoast}, A. and {Pei}, L. and {Penny}, Matthew T. and {Poleski}, Rados{\l}aw and {Rafter}, Stephen and {Romero-Colmenero}, E. and {Runnoe}, Jessie and {Sand}, David J. and {Schimoia}, Jaderson S. and {Sergeev}, S.~G. and {Shappee}, B.~J. and {Simonian}, Gregory V. and {Somers}, Garrett and {Spencer}, M. and {Starkey}, D.~A. and {Stevens}, Daniel J. and {Tayar}, Jamie and {Treu}, T. and {Valenti}, Stefano and {Van Saders}, J. and {Villanueva}, Jr., S. and {Villforth}, C. and {Weiss}, Yaniv and {Winkler}, H. and {Zhu}, W.},
        title = "{Reverberation Mapping of Optical Emission Lines in Five Active Galaxies}",
      journal = {\apj},
         year = 2017,
        month = may,
       volume = {840},
       number = {2},
          eid = {97},
        pages = {97},
          doi = {10.3847/1538-4357/aa6d52},
archivePrefix = {arXiv},
       eprint = {1610.00008},
 primaryClass = {astro-ph.GA},
       adsurl = {https://ui.adsabs.harvard.edu/abs/2017ApJ...840...97F}
}

@ARTICLE{Torralba2026b,
       author = {{Torralba}, Alberto and {Matthee}, Jorryt and {Pezzulli}, Gabriele and {Naidu}, Rohan P. and {Ishikawa}, Yuzo and {Brammer}, Gabriel B. and {Chang}, Seok-Jun and {Chisholm}, John and {de Graaff}, Anna and {D'Eugenio}, Francesco and {Di Cesare}, Claudia and {Eilers}, Anna-Christina and {Greene}, Jenny E. and {Gronke}, Max and {Iani}, Edoardo and {Kokorev}, Vasily and {Kotiwale}, Gauri and {Kramarenko}, Ivan and {Ma}, Yilun and {Mascia}, Sara and {Navarrete}, Benjam{\'\i}n and {Nelson}, Erica and {Oesch}, Pascal and {Simcoe}, Robert A. and {Wuyts}, Stijn},
        title = "{The warm outer layer of a little red dot as the source of [Fe II] and collisional Balmer lines with scattering wings}",
      journal = {\aap},
         year = 2026,
        month = feb,
       volume = {707},
          eid = {A75},
        pages = {A75},
          doi = {10.1051/0004-6361/202557537},
archivePrefix = {arXiv},
       eprint = {2510.00103},
 primaryClass = {astro-ph.GA},
       adsurl = {https://ui.adsabs.harvard.edu/abs/2026A&A...707A..75T}
}

@ARTICLE{Murray1995,
       author = {{Murray}, N. and {Chiang}, J. and {Grossman}, S.~A. and {Voit}, G.~M.},
        title = "{Accretion Disk Winds from Active Galactic Nuclei}",
      journal = {\apj},
         year = 1995,
        month = oct,
       volume = {451},
        pages = {498},
          doi = {10.1086/176238},
       adsurl = {https://ui.adsabs.harvard.edu/abs/1995ApJ...451..498M}
}

@ARTICLE{Trinca2026,
       author = {{Trinca}, Alessandro and {Lupi}, Alessandro and {Haardt}, Francesco and {Madau}, Piero},
        title = "{You can't see me: Super-Eddington growth hindering X-ray detection in high-z broad-line active galactic nuclei}",
      journal = {\aap},
         year = 2026,
        month = jun,
       volume = {710},
          eid = {A289},
        pages = {A289},
          doi = {10.1051/0004-6361/202659544},
       adsurl = {https://ui.adsabs.harvard.edu/abs/2026A&A...710A.289T}
}

@ARTICLE{Madau2026,
  author   = {{Madau}, Piero},
  title    = {Chasing the light: Shadowing, collimation, and the super-Eddington growth of infant black holes in JWST broad-line AGNs},
  journal  = {\aap},
  year     = {2026},
  volume   = {708},
  pages    = {A116},
  doi      = {10.1051/0004-6361/202659244},
  adsurl   = {https://ui.adsabs.harvard.edu/abs/2026A&A...708A.116M}
}

@ARTICLE{Maiolino2025,
       author = {{Maiolino}, Roberto and {Risaliti}, Guido and {Signorini}, Matilde and {Trefoloni}, Bartolomeo and {Juod{\v{z}}balis}, Ignas and {Scholtz}, Jan and {{\"U}bler}, Hannah and {D'Eugenio}, Francesco and {Carniani}, Stefano and {Fabian}, Andy and {Ji}, Xihan and {Mazzolari}, Giovanni and {Bertola}, Elena and {Brusa}, Marcella and {Bunker}, Andrew J. and {Charlot}, Stephane and {Comastri}, Andrea and {Cresci}, Giovanni and {DeCoursey}, Christa Noel and {Egami}, Eiichi and {Fiore}, Fabrizio and {Gilli}, Roberto and {Perna}, Michele and {Tacchella}, Sandro and {Venturi}, Giacomo},
        title = "{JWST meets Chandra: a large population of Compton thick, feedback-free, and intrinsically X-ray weak AGN, with a sprinkle of SNe}",
      journal = {\mnras},
         year = 2025,
        month = apr,
       volume = {538},
       number = {3},
        pages = {1921-1943},
          doi = {10.1093/mnras/staf359},
archivePrefix = {arXiv},
       eprint = {2405.00504},
 primaryClass = {astro-ph.GA},
       adsurl = {https://ui.adsabs.harvard.edu/abs/2025MNRAS.538.1921M}
}

@ARTICLE{MaiolinoAGN,
       author = {{Maiolino}, Roberto and {Scholtz}, Jan and {Curtis-Lake}, Emma and {Carniani}, Stefano and {Baker}, William and {de Graaff}, Anna and {Tacchella}, Sandro and {{\"U}bler}, Hannah and {D'Eugenio}, Francesco and {Witstok}, Joris and {Curti}, Mirko and {Arribas}, Santiago and {Bunker}, Andrew J. and {Charlot}, St{\'e}phane and {Chevallard}, Jacopo and {Eisenstein}, Daniel J. and {Egami}, Eiichi and {Ji}, Zhiyuan and {Jones}, Gareth C. and {Lyu}, Jianwei and {Rawle}, Tim and {Robertson}, Brant and {Rujopakarn}, Wiphu and {Perna}, Michele and {Sun}, Fengwu and {Venturi}, Giacomo and {Williams}, Christina C. and {Willott}, Chris},
        title = "{JADES: The diverse population of infant black holes at 4 < z < 11: Merging, tiny, poor, but mighty}",
      journal = {\aap},
         year = 2024,
        month = nov,
       volume = {691},
          eid = {A145},
        pages = {A145},
          doi = {10.1051/0004-6361/202347640},
       adsurl = {https://ui.adsabs.harvard.edu/abs/2024A&A...691A.145M}
}

@ARTICLE{Matthee2024,
       author = {{Matthee}, Jorryt and {Naidu}, Rohan P. and {Brammer}, Gabriel and {Chisholm}, John and {Eilers}, Anna-Christina and {Goulding}, Andy and {Greene}, Jenny and {Kashino}, Daichi and {Labbe}, Ivo and {Lilly}, Simon J. and {Mackenzie}, Ruari and {Oesch}, Pascal A. and {Weibel}, Andrea and {Wuyts}, Stijn and {Xiao}, Mengyuan and {Bordoloi}, Rongmon and {Bouwens}, Rychard and {van Dokkum}, Pieter and {Illingworth}, Garth and {Kramarenko}, Ivan and {Maseda}, Michael V. and {Mason}, Charlotte and {Meyer}, Romain A. and {Nelson}, Erica J. and {Reddy}, Naveen A. and {Shivaei}, Irene and {Simcoe}, Robert A. and {Yue}, Minghao},
        title = "{Little Red Dots: An Abundant Population of Faint Active Galactic Nuclei at z {\ensuremath{\sim}} 5 Revealed by the EIGER and FRESCO JWST Surveys}",
      journal = {\apj},
         year = 2024,
        month = mar,
       volume = {963},
       number = {2},
          eid = {129},
        pages = {129},
          doi = {10.3847/1538-4357/ad2345},
archivePrefix = {arXiv},
       eprint = {2306.05448},
 primaryClass = {astro-ph.GA},
       adsurl = {https://ui.adsabs.harvard.edu/abs/2024ApJ...963..129M}
}

@INCOLLECTION{Peterson2006,
       author = {{Peterson}, B.~M.},
        title = "{The Broad-Line Region in Active Galactic Nuclei}",
    booktitle = {Physics of Active Galactic Nuclei at all Scales},
         year = 2006,
       editor = {{Alloin}, Danielle},
       volume = {693},
        pages = {77},
          doi = {10.1007/3-540-34621-X_3},
       adsurl = {https://ui.adsabs.harvard.edu/abs/2006LNP...693...77P}
}

@ARTICLE{Taylor2025_BHMF,
       author = {{Taylor}, Anthony J. and {Finkelstein}, Steven L. and {Kocevski}, Dale D. and {Jeon}, Junehyoung and {Bromm}, Volker and {Amor{\'\i}n}, Ricardo O. and {Arrabal Haro}, Pablo and {Backhaus}, Bren E. and {Bagley}, Micaela B. and {Banados}, Eduardo and {Bhatawdekar}, Rachana and {Brooks}, Madisyn and {Calabr{\`o}}, Antonello and {Ch{\'a}vez Ortiz}, {\'O}scar A. and {Cheng}, Yingjie and {Cleri}, Nikko J. and {Cole}, Justin W. and {Davis}, Kelcey and {Dickinson}, Mark and {Donnan}, Callum and {Dunlop}, James S. and {Ellis}, Richard S. and {Fern{\'a}ndez}, Vital and {Fontana}, Adriano and {Fujimoto}, Seiji and {Giavalisco}, Mauro and {Grazian}, Andrea and {Guo}, Jingsong and {Hathi}, Nimish P. and {Holwerda}, Benne W. and {Hirschmann}, Michaela and {Inayoshi}, Kohei and {Kartaltepe}, Jeyhan S. and {Khusanova}, Yana and {Koekemoer}, Anton M. and {Kokorev}, Vasily and {Larson}, Rebecca L. and {Leung}, Gene C.~K. and {Lucas}, Ray A. and {McLeod}, Derek J. and {Napolitano}, Lorenzo and {Onoue}, Masafusa and {Pacucci}, Fabio and {Papovich}, Casey and {P{\'e}rez-Gonz{\'a}lez}, Pablo G. and {Pirzkal}, Nor and {Somerville}, Rachel S. and {Trump}, Jonathan R. and {Wilkins}, Stephen M. and {Yung}, L.~Y. Aaron and {Zhang}, Haowen},
        title = "{Broad-line AGNs at 3.5 < z < 6: The Black Hole Mass Function and a Connection with Little Red Dots}",
      journal = {\apj},
         year = 2025,
        month = jun,
       volume = {986},
       number = {2},
          eid = {165},
        pages = {165},
          doi = {10.3847/1538-4357/add15b},
archivePrefix = {arXiv},
       eprint = {2409.06772},
 primaryClass = {astro-ph.GA},
       adsurl = {https://ui.adsabs.harvard.edu/abs/2025ApJ...986..165T}
}

@ARTICLE{Zucchi2026,
       author = {{Zucchi}, Greta and {Ji}, Xihan and {Madau}, Piero and {Maiolino}, Roberto and {Juod{\v{z}}balis}, Ignas and {D'Eugenio}, Francesco and {Geris}, Sophia and {Isobe}, Yuki},
        title = "{Black holes in the shadows: The missing high-ionization lines in the earliest JWST active galactic nuclei}",
      journal = {\aap},
         year = 2026,
        month = feb,
       volume = {707},
          eid = {A52},
        pages = {A52},
          doi = {10.1051/0004-6361/202557687},
archivePrefix = {arXiv},
       eprint = {2510.10772},
 primaryClass = {astro-ph.GA},
       adsurl = {https://ui.adsabs.harvard.edu/abs/2026A&A...707A..52Z}
}

@ARTICLE{Ubler2023,
       author = {{{\"U}bler}, Hannah and {Maiolino}, Roberto and {Curtis-Lake}, Emma and {P{\'e}rez-Gonz{\'a}lez}, Pablo G. and {Curti}, Mirko and {Perna}, Michele and {Arribas}, Santiago and {Charlot}, St{\'e}phane and {Marshall}, Madeline A. and {D'Eugenio}, Francesco and {Scholtz}, Jan and {Bunker}, Andrew and {Carniani}, Stefano and {Ferruit}, Pierre and {Jakobsen}, Peter and {Rix}, Hans-Walter and {Rodr{\'\i}guez Del Pino}, Bruno and {Willott}, Chris J. and {Boeker}, Torsten and {Cresci}, Giovanni and {Jones}, Gareth C. and {Kumari}, Nimisha and {Rawle}, Tim},
        title = "{GA-NIFS: A massive black hole in a low-metallicity AGN at z {\ensuremath{\sim}} 5.55 revealed by JWST/NIRSpec IFS}",
      journal = {\aap},
         year = 2023,
        month = sep,
       volume = {677},
          eid = {A145},
        pages = {A145},
          doi = {10.1051/0004-6361/202346137},
archivePrefix = {arXiv},
       eprint = {2302.06647},
 primaryClass = {astro-ph.GA},
       adsurl = {https://ui.adsabs.harvard.edu/abs/2023A&A...677A.145U}
}

@ARTICLE{Wang2014,
       author = {{Wang}, Jian-Min and {Qiu}, Jie and {Du}, Pu and {Ho}, Luis C.},
        title = "{Self-shadowing Effects of Slim Accretion Disks in Active Galactic Nuclei: The Diverse Appearance of the Broad-line Region}",
      journal = {\apj},
         year = 2014,
        month = dec,
       volume = {797},
       number = {1},
          eid = {65},
        pages = {65},
          doi = {10.1088/0004-637X/797/1/65},
archivePrefix = {arXiv},
       eprint = {1410.5285},
 primaryClass = {astro-ph.GA},
       adsurl = {https://ui.adsabs.harvard.edu/abs/2014ApJ...797...65W}
}

@ARTICLE{Wang2025,
       author = {{Wang}, Bingjie and {de Graaff}, Anna and {Davies}, Rebecca L. and {Greene}, Jenny E. and {Leja}, Joel and {Brammer}, Gabriel B. and {Goulding}, Andy D. and {Miller}, Tim B. and {Suess}, Katherine A. and {Weibel}, Andrea and {Williams}, Christina C. and {Bezanson}, Rachel and {Boogaard}, Leindert A. and {Cleri}, Nikko J. and {Hirschmann}, Michaela and {Katz}, Harley and {Labb{\'e}}, Ivo and {Maseda}, Michael V. and {Matthee}, Jorryt and {McConachie}, Ian and {Naidu}, Rohan P. and {Oesch}, Pascal A. and {Rix}, Hans-Walter and {Setton}, David J. and {Whitaker}, Katherine E.},
        title = "{RUBIES: JWST/NIRSpec Confirmation of an Infrared-luminous, Broad-line Little Red Dot with an Ionized Outflow}",
      journal = {\apj},
         year = 2025,
        month = may,
       volume = {984},
       number = {2},
          eid = {121},
        pages = {121},
          doi = {10.3847/1538-4357/adc1ca},
archivePrefix = {arXiv},
       eprint = {2403.02304},
 primaryClass = {astro-ph.GA},
       adsurl = {https://ui.adsabs.harvard.edu/abs/2025ApJ...984..121W}
}

@ARTICLE{Proga2004,
       author = {{Proga}, Daniel and {Kallman}, Timothy R.},
        title = "{Dynamics of Line-driven Disk Winds in Active Galactic Nuclei. II. Effects of Disk Radiation}",
      journal = {\apj},
         year = 2004,
        month = dec,
       volume = {616},
       number = {2},
        pages = {688-695},
          doi = {10.1086/425117},
archivePrefix = {arXiv},
       eprint = {astro-ph/0408293},
 primaryClass = {astro-ph},
       adsurl = {https://ui.adsabs.harvard.edu/abs/2004ApJ...616..688P}
}

\end{document}